\documentclass[a4paper,11pt]{article}
\pdfoutput=1 

\usepackage{jinstpub} 

\usepackage{subcaption}                     
\usepackage{threeparttable}                 
\usepackage{tablefootnote}                  
\usepackage[sort&compress,numbers]{natbib}  
\usepackage[section]{placeins}              
\usepackage{cleveref}                       

\usepackage{lineno}
\modulolinenumbers[5]

\title{\boldmath Suppression of intrinsic neutron background in the Multi-Grid detector}

\author[a,b,c,1]{E.~Dian,\note{Corresponding author.}}
\author[b]{K.~Kanaki,}
\author[b]{A.~Khaplanov,}
\author[b]{T.~Kittelmann,}
\author[a,c]{P.~Zagyvai,}
\author[b,d]{R.~Hall-Wilton}

\affiliation[a]{Hungarian Academy of Sciences, Centre for Energy Research,\\
  29-33 Konkoly Thege Mikl\'{o}s street, 1121 Budapest, Hungary}
\affiliation[b]{European Spallation Source ESS ERIC, \\
  P.O Box 176, SE-221 00 Lund, Sweden}
\affiliation[c]{Budapest University of Technology and Economics, Institute of Nuclear Techniques, \\
   M\H uegyetem rakpart 9., 1111 Budapes, Hungary}
\affiliation[d]{Mid-Sweden University, \\
  SE-851 70 Sundsvall, Sweden}
\emailAdd{dian.eszter@energia.mta.hu}

\abstract{ 

  One of the key requirements for neutron scattering instruments is the  Signal-to-Background ratio (SBR). This is as well a design driving requirement for many instruments at the European Spallation Source (ESS), which aspires to be the brightest neutron source of the world. The SBR can be effectively improved with background reduction.
  The Multi-Grid, a large-area thermal neutron detector with a solid boron carbide converter, is a novel solution for chopper spectrometers. This detector 
  will be installed for the three prospective chopper spectrometers at the ESS.
  As the Multi-Grid detector is a large area detector with a complex structure, its intrinsic background and its suppression via advanced shielding design should be investigated in its complexity, as it cannot be naively calculated.   
 The intrinsic scattered neutron background and its effect on the SBR is determined via a detailed Monte Carlo simulation for the Multi-Grid detector module, designed for the CSPEC instrument at the ESS.
 The impact of  the detector vessel and the 
 neutron entrance window on scattering is determined, revealing the importance of an optimised internal detector shielding. The background-reducing capacity of common shielding geometries, like side-shielding and end-shielding is determined by using 
 perfect absorber as shielding material, and common shielding materials, like B$_{4}$C and Cd are also tested. 

 On the basis of the comparison of the effectiveness of the different shielding topologies and materials, recommendations are given for a combined shielding of the Multi-Grid detector module, optimised for increased SBR.}


\arxivnumber{arxiv to fill} 

\begin{document}

\maketitle
\flushbottom


\section{Introduction}\label{sec:intro}

The ongoing construction of the ESS~\cite{peggs2012,peggs2013,garoby2018} and the recent restructurisation of the $^{3}$He market~\cite{shea} have led to a significant investment in neutron detector development in the past eight~years. One of the main research fields is the development of a cost-effective, large area detector design for chopper spectroscopy, as a competitive alternative for the currently used $^{3}$He-tubes~\cite{zeitelhack2012}.  
Although an adequate Signal-to-Background Ratio (SBR) is a key requirement for neutron scattering instruments~\cite{cherkashyna_2014}, the inelastic instruments are particularly background-sensitive, therefore the background suppression should be one of the main focuses in the detector design. Since the chopper spectrometers are 
equipped with large area -- and therefore mostly large volume -- detectors, the intrinsic background, induced within the detector by neutrons, should also be considered. 
As the detection efficiency for thermal neutrons is much higher than for other typical sources of background, like fast neutrons and gamma-radiation~\cite{khaplanov2013a, messi2018}, and neutron induced intrinsic gamma background has already been studied and found to be negligible~\cite{dian2017}, thermal neutron scattering is likely to be the dominant source of background in cold and thermal instruments, and is therefore 
the focus of the current study.

The different sources of scattered neutron background have already been identified in a potent $^{3}$He-substitute demonstrator 
for chopper spectroscopy~\cite{dian2018}. The detector was tested at the CNCS instrument at SNS~\cite{khaplanov2017}, and a realistic model of the detector and the sample environment was implemented in Geant4~\cite{geant4, agostinelli2003,allison2006,allison2016b}. 
The model was validated 
against measured data, 
scattered neutron background 
reproduced, and instrument and detector related background components were distinguished, like scattering on the sample environment, on the tank gas or in the detector~\cite{dian2018}. The study revealed that the impact of the intrinsic detector background is comparable  with the ones of instrument-related sources.
%
This reemphasises that the study and optimisation of the internal detector shielding is needed.

The most potent novel detector design for chopper spectrometers is the Multi-Grid detector, invented at the Institute Laue-Langevin (ILL)~\cite{ill,ill_patent}, and jointly developed at ILL and the ESS~\cite{mg_patent_ill, andersen_2012, khaplanov2013b}, as this detector has been chosen to serve the three chopper spectrometers at the ESS, the CSPEC~\cite{CSPECprop}, T-REX~\cite{TREXprop} and VOR~\cite{deen2015} instruments. 
It is an Ar/CO$_{2}$-filled proportional chamber with a solid boron carbide converter, enriched in $^{10}$B. The few $\mu$m thick converter layers are 
supported on an aluminium grid, 
designating the sensitive volume of the detector, and each detector module is placed in a few mm thick aluminium housing.  
The source of the intrinsic scattered neutron background is this aluminium structure; 
because of its complexity, multiple scattering volumes and surfaces have to be considered as source of background, 
but it also leaves room for effective background reduction via the application of a targeted, multi-element shielding.

Thanks to the great effort that has been invested in the development of neutron scattering simulation tools~\cite{kanaki2018_inpresscorrectedproof} in the past few years, this otherwise hard-to-study intrinsic detector background 
can now be investigated in detail via realistic Monte Carlo simulations.

The aluminium-sourced intrinsic background now can be simulated with the usage of NXSG4~\cite{nxsg4, kittelmann2015b} and NCrystal~\cite{ncrystal} tools, that allow to model neutron interaction with crystalline materials, 
including both Bragg diffraction and inelastic/incoherent processes, in Geant4~\cite{geant4, agostinelli2003,allison2006} and McStas~\cite{mcstas1,mcstas2}. 
These and other recently developed neutron simulation tools, like the MCPL~\cite{kittelmann2017,mcpl,mcpl2}, a file format to store and transfer full particle state information between different Monte Carlo codes, are already incorporated and ready to
use in the 
supportive, user friendly ESS Coding Framework~\cite{dgcodechep2013}, developed in the ESS Detector Group, facilitating the tailoring of the detector designs to the scientific application.  


The aim of the current study is to increase the 
SBR in the Multi-Grid detector module of the CSPEC instrument, by optimising the shielding to reduce intrinsic scattered neutron background. For this purpose, detailed Monte Carlo simulations are performed with a recently developed and validated Geant4-based Multi-Grid model~\cite{dian2018}, tailored to the current 
CSPEC detector design.
The implemented detector model and the simulation setup are described in Section~\ref{model}. 
The comparison of different coating designs is presented on Section~\ref{coat}, while the impact of neutron scattering on the aluminium housing, especially on the vessel window is studied and discussed in Section~\ref{win}. To optimise the shielding, an ideal total absorber and common shielding materials, like B$_{4}$C and Cd are applied in typical shielding geometries. 
This way the background-reducing capability of shielding applied at different locations is determined (see Section~\ref{black}), and then the sepective impacts of commonly used shielding materials are compared for each geometry (Section~\ref{sh}).

The obtained results are 
discussed in 
the conclusion, where this holistic shielding optimisation approach and Multi-Grid specific shielding recommendations are both presented.







\FloatBarrier

\section{Simulation of the Multi-Grid detector modules}\label{model}

The Monte Carlo modelling of the Multi-Grid detector has been going on in the past couple of years. A detailed, realistic and parameterised model of the detector was developed~\cite{dian2018} in Geant4, within the afore-mentioned ESS Coding Framework~\cite{dgcodechep2013}. 
%
Two detectors that have been previously built and tested in neutron scattering instruments, IN6~\cite{khaplanov2014} and CNCS~\cite{khaplanov2017} at the ILL and SNS neutron sources, respectively, were simulated, and the model was validated via qualitative and quantitative reproduction of measured data. 
This validated model is adapted for the current detector design of the CSPEC instrument~\cite{khaplanov2018_psnd}. The simulation setup involves an isotropic point source and a single detector module.

\FloatBarrier

\subsection{Geant4 model of CSPEC detector module}\label{module}

The Multi-Grid detector is an Ar/CO$_{2}$-filled proportional chamber, consists 
of the so-called `grids': aluminium frames divided into $2.5 \times 2.5 \times 1$~cm$^{3}$ cells by 0.5~mm thick aluminium `blades', coated on 
both sides with enriched $^{10}$B$_{4}$C, 97\% enriched in $^{10}$B~\cite{hoglund2012a, c.hoglund2015b, LiU2016b}. 
%
%
%
The coating thickness on the so-called `short blades', perpendicular to the incident  neutron beam, varies between 0.5-1.5~$\mu$m through the depth of the grid, to increase the efficiency. In addition, unlike the previous 
demonstrators, in the CSPEC detector the `long blades', parallel with the incident neutron beam, are also coated on 
both sides with an identical 1.0~$\mu$m thickness. 

\begin{figure}[ht!]
  \centering
    \begin{subfigure}[c]{0.62\textwidth}
      \includegraphics[width=\textwidth]{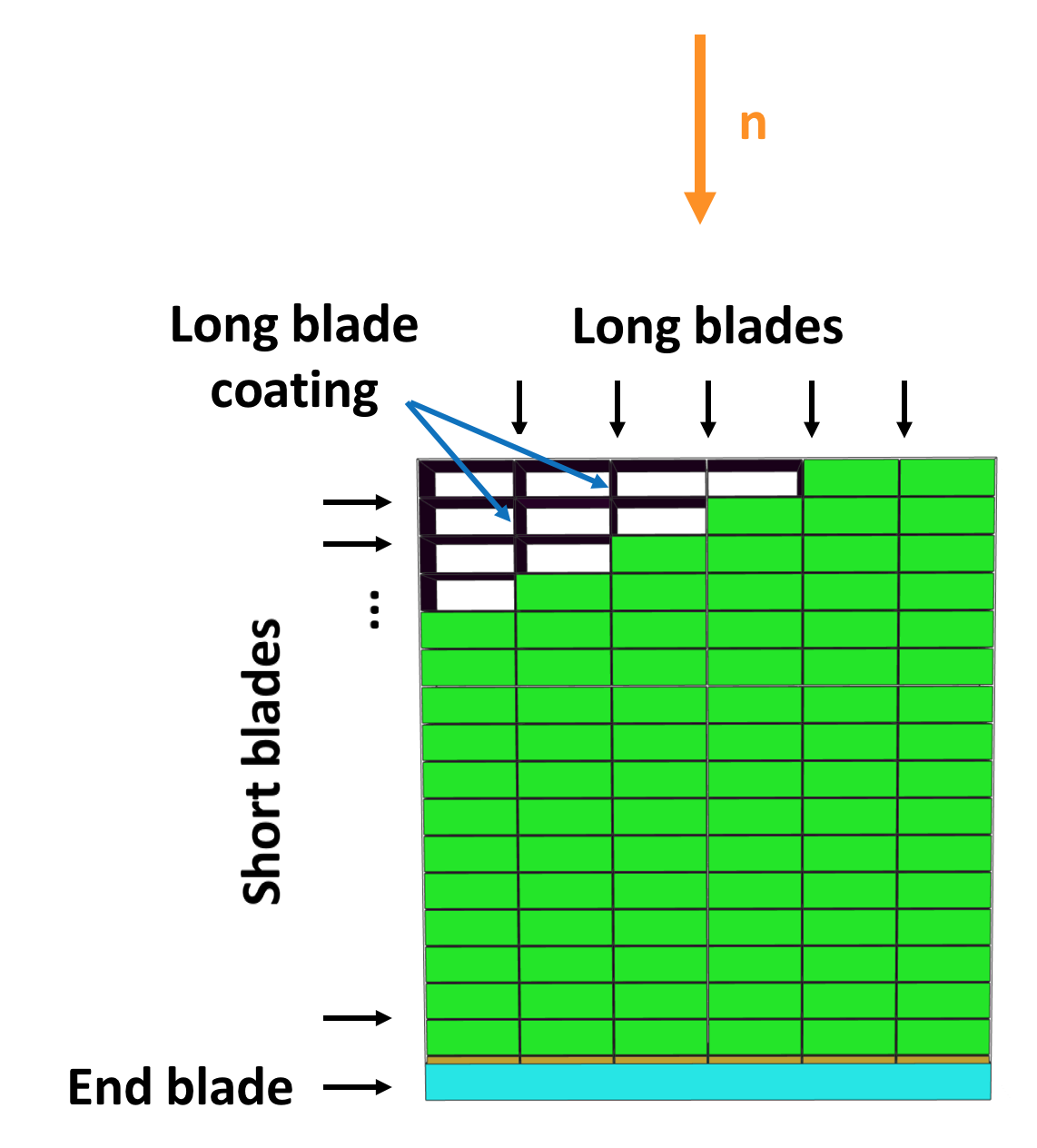}
      \caption{ \label{grid}}
    \end{subfigure}
    \begin{subfigure}[c]{0.28\textwidth}
      \includegraphics[width=\textwidth]{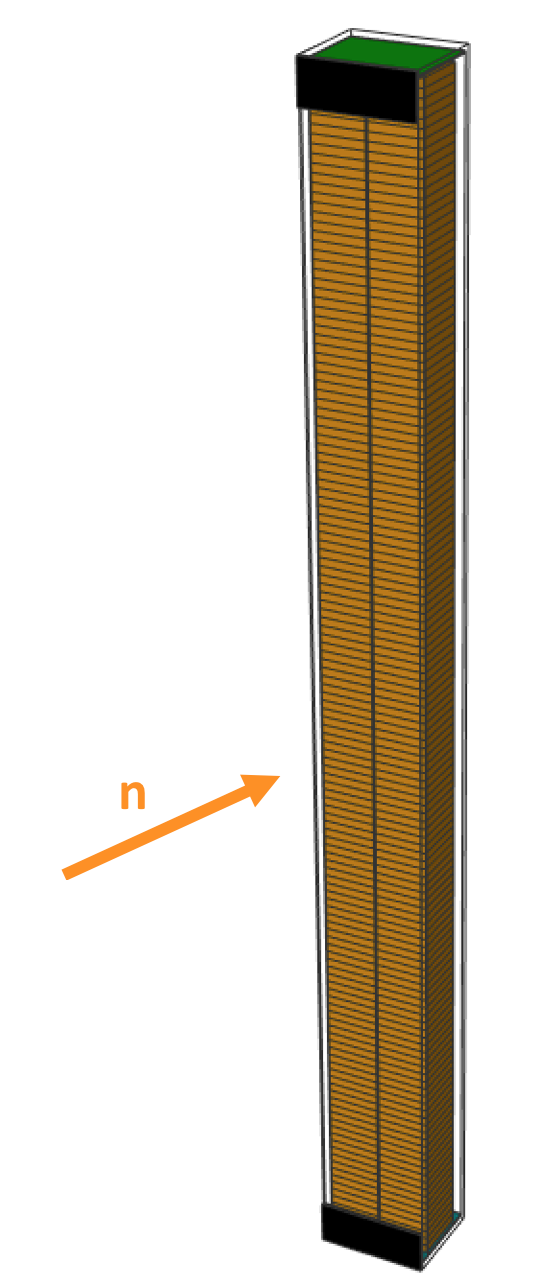}
      \caption{ \label{baregrid}}
    \end{subfigure}

  \caption{\footnotesize The CSPEC Multi-Grid detector module geometry implemented in Geant4. Single simulated grid from top view~(\subref{grid}) and whole detector module~(\subref{baregrid}) with front shielding in a transparent vessel. In~\subref{grid} green represents the counting gas, cyan is the rear aluminium blade and brown is shielding. In~\subref{baregrid} green is the PCB, black is the front shielding and brown is the aluminium grid. \label{g4geom} }  
\end{figure}

The grids are 6 cells wide and 16 cells deep, leading to 32 converter layers on the short blades, in order to meet the efficiency of $^{3}$H-tubes.
%
These grids 
are stacked, forming 3.5~m high columns. The anode wires go through the length of the columns in the  6~$\times$~16  channels formed by the cells in each grid, and the grids serve as cathodes. 
A realistic, one-to-one model of the detector module is implemented for the Geant4 simulation. The constructed 
model is shown in figure~\ref{g4geom}, and the used parameter set is given in Table~\ref{tab:geomparam}.

\FloatBarrier

The implemented model consists of the 2 columns of grids placed in an aluminium vessel filled with Ar/CO$_{2}$. Minor simplifications are applied for the aluminium vessel: it 
is 
a rectangular parallelepiped, with a flat front window, unlike the one to be built, which is designed to be slightly curved, so multiple modules can fit together. A further simplification is that the printed circuit boards (PCB) of read-out electronics placed in the detector vessel are represented as layers of aluminium and polyethylene at the top and the bottom of the vessel. 
Also, 2 sheets of 
shielding are applied 
at the top and bottom of the front window, 
adequately sized to shield the PCBs from neutrons arriving from the sample position, as it is planned for the real detector (see  figure~\ref{baregrid}). 



\begin{table}[htbp]
  \centering
  \caption{Major default parameters of CSPEC Multi-Grid detector module, grouped as geometrical parameters, processes and materials.}
  \smallskip
  \label{tab:geomparam}
  \begin{tabular}{|ll|c|}
    \hline
    Number of cells               & width~(x)        &   6                  \\
                                  & depth~(z)        &  16                  \\
    Number of grids in columns    &                  & 140                  \\
    Number of columns             &                  &   2                  \\
    Cell size~                    & width~(x)        &   2.5~cm             \\
                                  & height~(y)       &   2.4~cm             \\
                                  & depth~(z)        &   0.95~cm             \\
    Coating thickness             & short blade      &   0.5-1.5~$\mu$m      \\
                                  & (parallel with window)     &             \\
                                  & long blade       &   1.0~$\mu$m          \\
                                  & (perpendicular to window)  &             \\
    Frame thickness               &                  &   0.5~mm             \\
    Frame end thickness           &                  &  10.0~mm             \\
    Blade thickness               & short blade      &   0.5~mm             \\
                                  & long blade       &   0.5~mm            \\                       

    Vessel window thickness       &                  &   4.0~mm              \\
    Vessel wall thickness         &                  &   4.0~mm              \\
    Sample-detector front face distance      &       &  3.5~m      \\ 
    \hline
    Geant4 Physics list           &                  & ESS\_QGSP\_BIC\_HP\_TS~\cite{dgcodechep2013}+NCrystal               \\
    \hline
    Vessel material               &                  &  Al                  \\
    PCB material                  &                  &  Al, polyethylene    \\
    Frame material                &                  &  Al                  \\
    Counting gas                  &                  &  Ar/CO$_{2}$         \\
                                  &                  &  80/20~\% volume           \\
    Coating                       &                  & $^{10}$B$_{4}$C       \\
                                  &                  & Boron is 97~\% enriched in $^{10}$B      \\
        
    \hline

  \end{tabular}
\end{table}

\begin{figure}[ht!]
  \centering  
    \begin{subfigure}[b]{0.66\textwidth}
      \includegraphics[width=\textwidth]{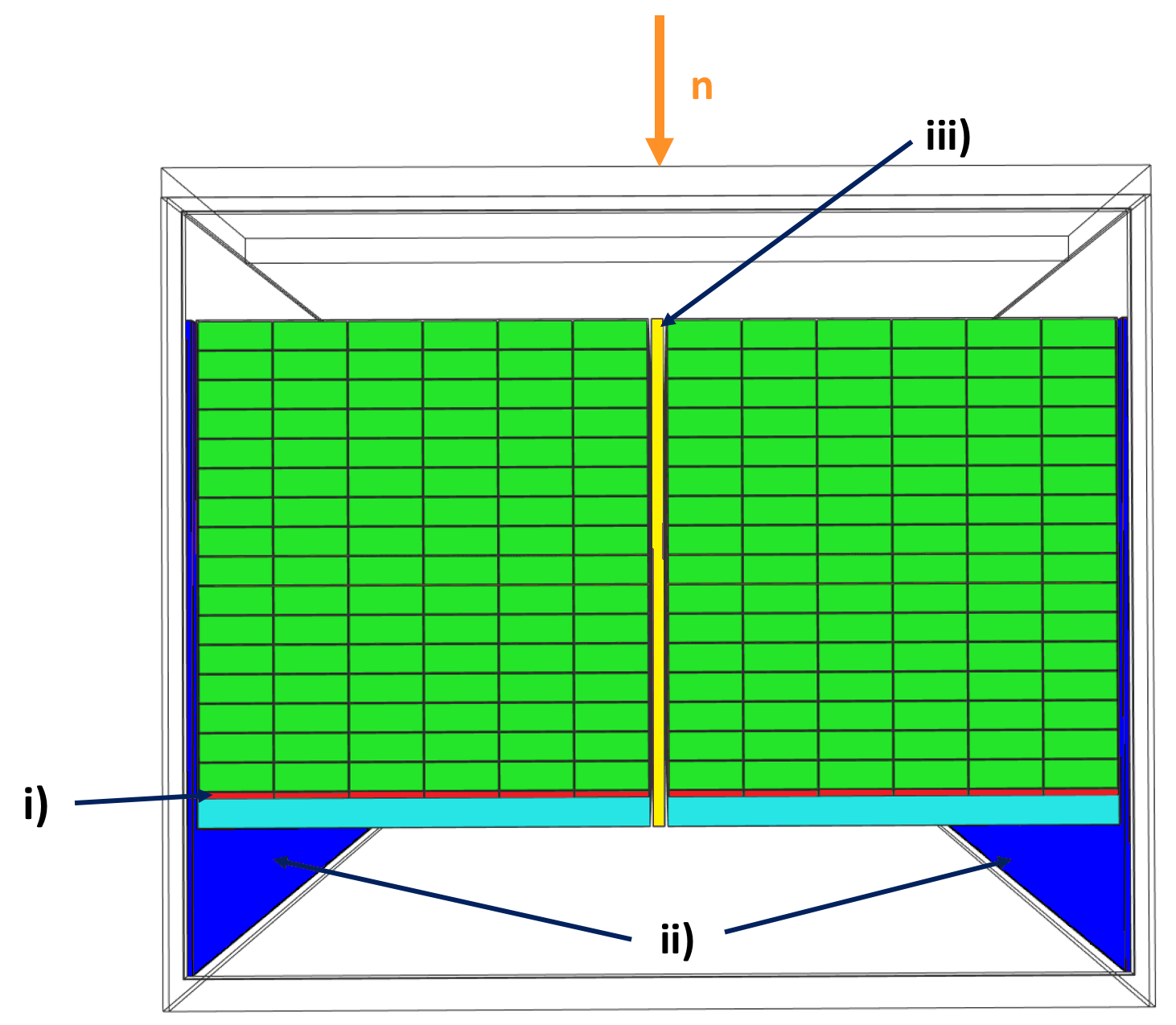}
      \caption{\label{dettop}}
    \end{subfigure}
    \begin{subfigure}[b]{0.33\textwidth}
      \includegraphics[width=\textwidth]{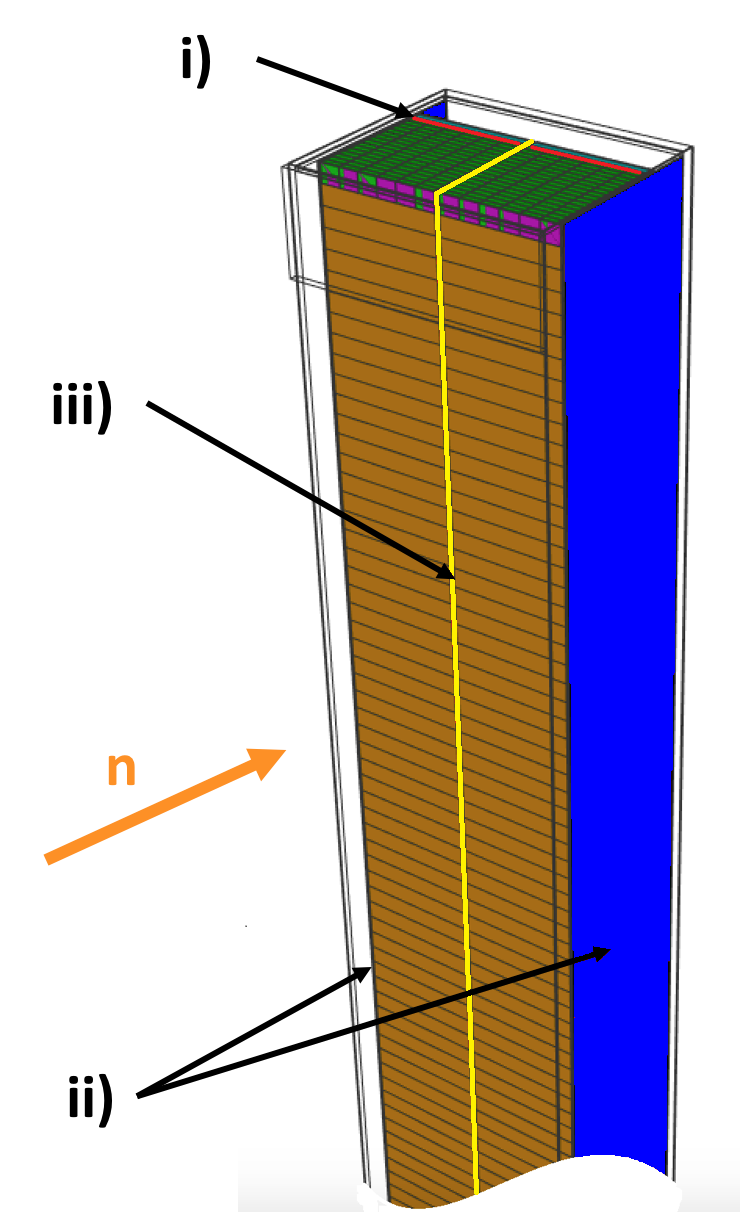}
      \caption{\label{detside}}
    \end{subfigure}

  \caption{\footnotesize Shielded CSPEC Multi-Grid detector module geometry. Top view~(\subref{dettop}) and side view~(\subref{detside}) with the studied shielding topologies marked with: red for \textit{i)}~`End-shielding', blue for \textit{ii)}~`Side-shielding' and yellow for \textit{iii)}~`Interstack-shielding'. Counting gas is shown in green, the grid is brown with cyan rear blade, and the incident neutron beam is indicated in orange. \label{g4sh} }  
\end{figure} 

 

The detector model involves pre-defined volumes for shielding materials in the most common places of the detector. The size of all shielding volumes are maximised insofar it is possible while keeping the dead area to the minimum.  
In this study, 
three shielding geometries 
are compared (see figure~\ref{g4sh}):

\begin{itemize}
\item `End-shielding': Layers of shielding~(see figure~\ref{g4sh}, \textit{i)}, red) applied in each grid, placed between the last row of cells~(green) and the 1~cm thick aluminium rear blade~(cyan) of the grid, to prevent backscattering from the latter. The surface of the shielding meets the dimensions of the cell. The maximum feasible thickness is 2~mm, defined by the space 
  between the last coated blade and the end blade.
\item `Side-shielding': Layers of shielding~(see figure~\ref{g4sh}, \textit{ii)}, blue) applied on the inner side of the vessel wall (see figure~\ref{detside}, transparent), as 
  \cite{dian2018} indicated the importance of scattering on the vessel. The shielding surface is defined by the size of the vessel wall. 
  The shielding sheets do not extend beyond the front face of the columns, as this would interfere with the neighbouring module placement. The maximum feasible thickness is 3.5~mm, i.e.\ the gap between the columns and the vessel wall. 
\item `Interstack-shielding': A sheet of shielding~(see figure~\ref{g4sh}, \textit{iii)}, yellow) placed between the two columns of grids~(see figure~\ref{detside}, brown), to prevent cross-talk. The shielding surface area meets the dimensions of the columns, and the maximum feasible thickness is 6~mm, i.e. the width of the gap between the columns.
\end{itemize}

The listed shielding topologies are modelled with both `black material' (ideal total absorber) 
and common shielding materials. All shielding materials are used  with natural isotope composition and in a realistic chemical form, with a representative carrier matrix, if necessary: 

\begin{itemize}
\item  $\rm B_{4}C$
\item  Cd 
\item  LiF
\item $\rm 50\% \  Gd_{2}O_{3}$ + 50\% polyethylene (representing acrylic paint as a typical carrier)
\item black material, i.e.\ perfectly opaque to incident neutrons
\end{itemize}

All materials in the model are 
the compositions of standard Geant4 materials except aluminium, 
whose poly-crystalline structure is enabled with the help of the NCrystal library.
The black material is emulated via an MCPL particle filter, which is set to kill all particles that enter the respective volumes. A customised physics list is used for the simulations due to the thermal scattering on the high hydrogen-content of the polyethylene in the PCBs. 

The neutron scattering in idealistic constructions, e.g.~with black shielding material, or in the absence of certain components or material, like the aluminium  window, can only be studied in simulation.


\subsection{Description of configuration and simulated quantities}\label{config}

\begin{figure}[ht!]
  \centering
    \includegraphics[width=0.63\textwidth]{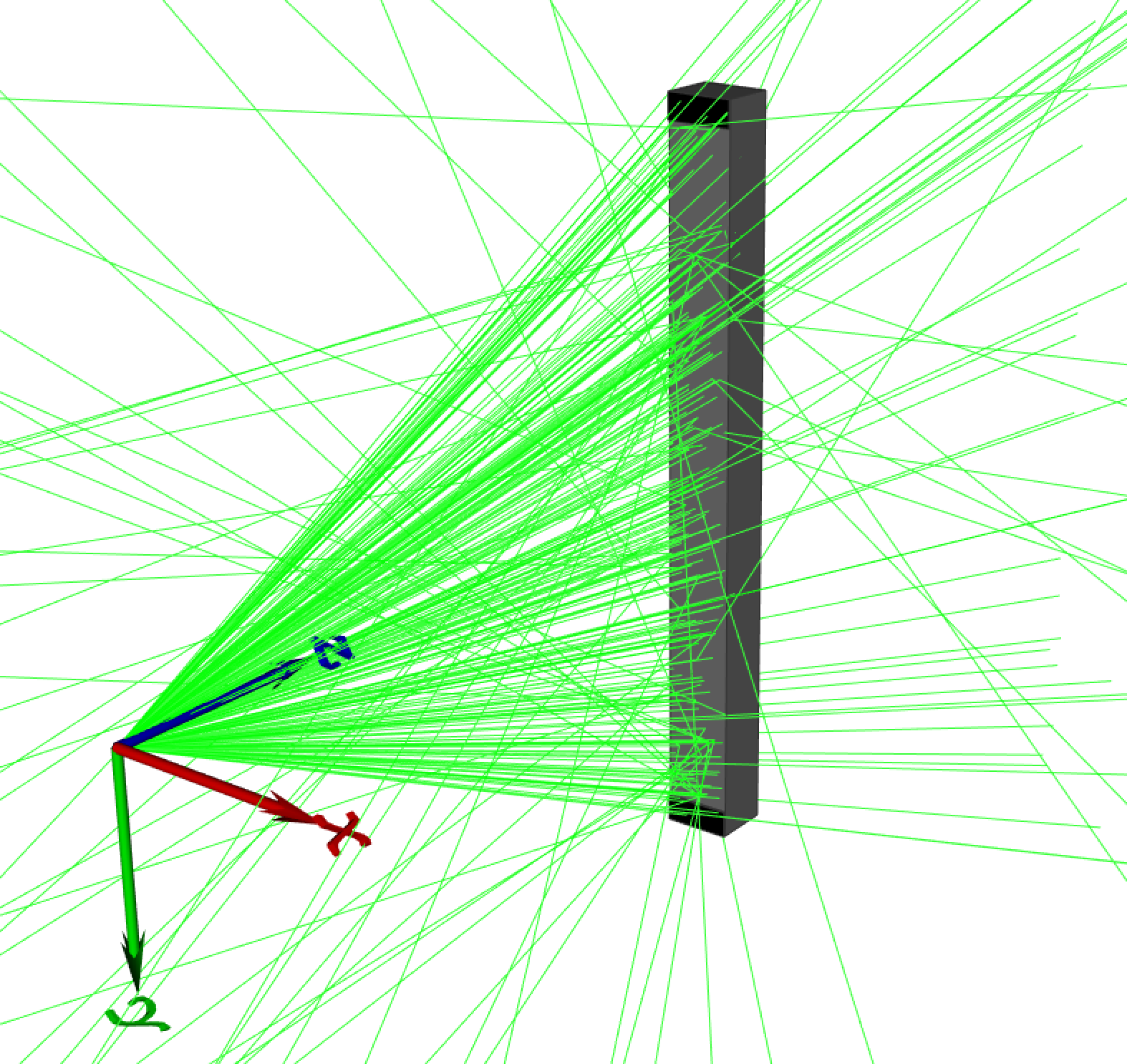}
  \caption{\footnotesize Geometry view of CSPEC Geant4 model with isotropic point source of neutrons (in green), targeted towards the detector window.  \label{g4source} }  
\end{figure}

In order to get a clear view of the intrinsic scattering, the detector is irradiated with mono-energetic neutrons, and all instrument related effects are excluded from the Geant4 simulation.  
The neutrons are generated isotropically at the sample position as a point source and are targeting the detector window, as shown in figure~\ref{g4source}.
The distance from the source to the detector front window is 3.5~m, and the sensitive area of the detector window covers 0.080~sr solid angle.
The neutron wavelengths are chosen at 0.4, 1, 1.8, 4.0 and 10.0~\AA, 
covering the operational range of the CSPEC instrument, extended with 
  neutrons down to the Cd/Gd cutoff. All simulations are performed with $2 \times 10^{7}$ neutrons.

\FloatBarrier


The effectiveness of the different shielding geometries and materials is studied by their impact on the SBR. For this comparison, a `reference' is defined: the detector in the vessel, with long blade coating, but without any shielding. 

\begin{figure}[ht!]
   \centering
  \begin{subfigure}[b]{0.49\textwidth}
    \includegraphics[width=\textwidth]{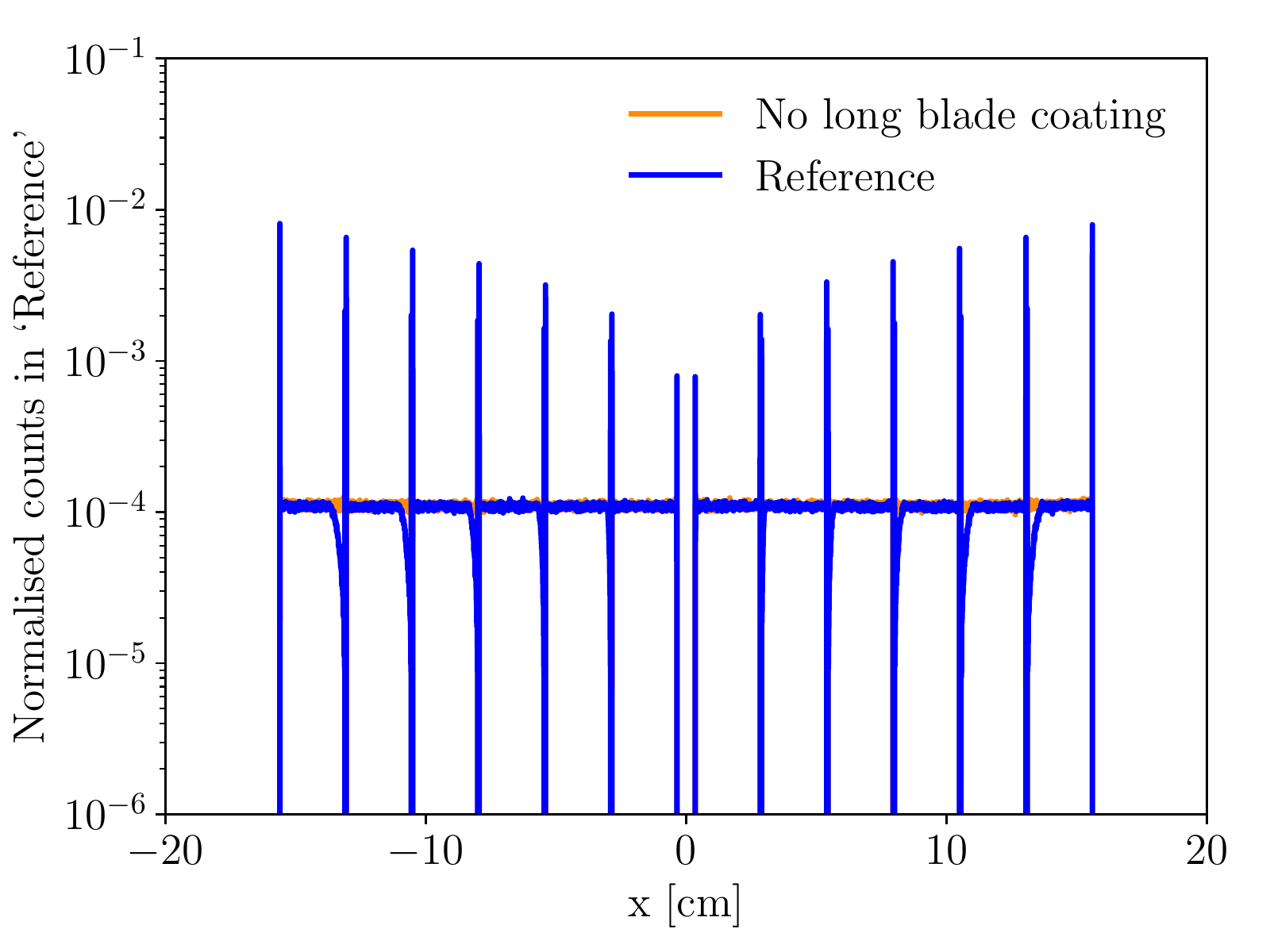}
    \caption{\label{hitxAll}}
  \end{subfigure}
  \begin{subfigure}[b]{0.49\textwidth}
    \includegraphics[width=\textwidth]{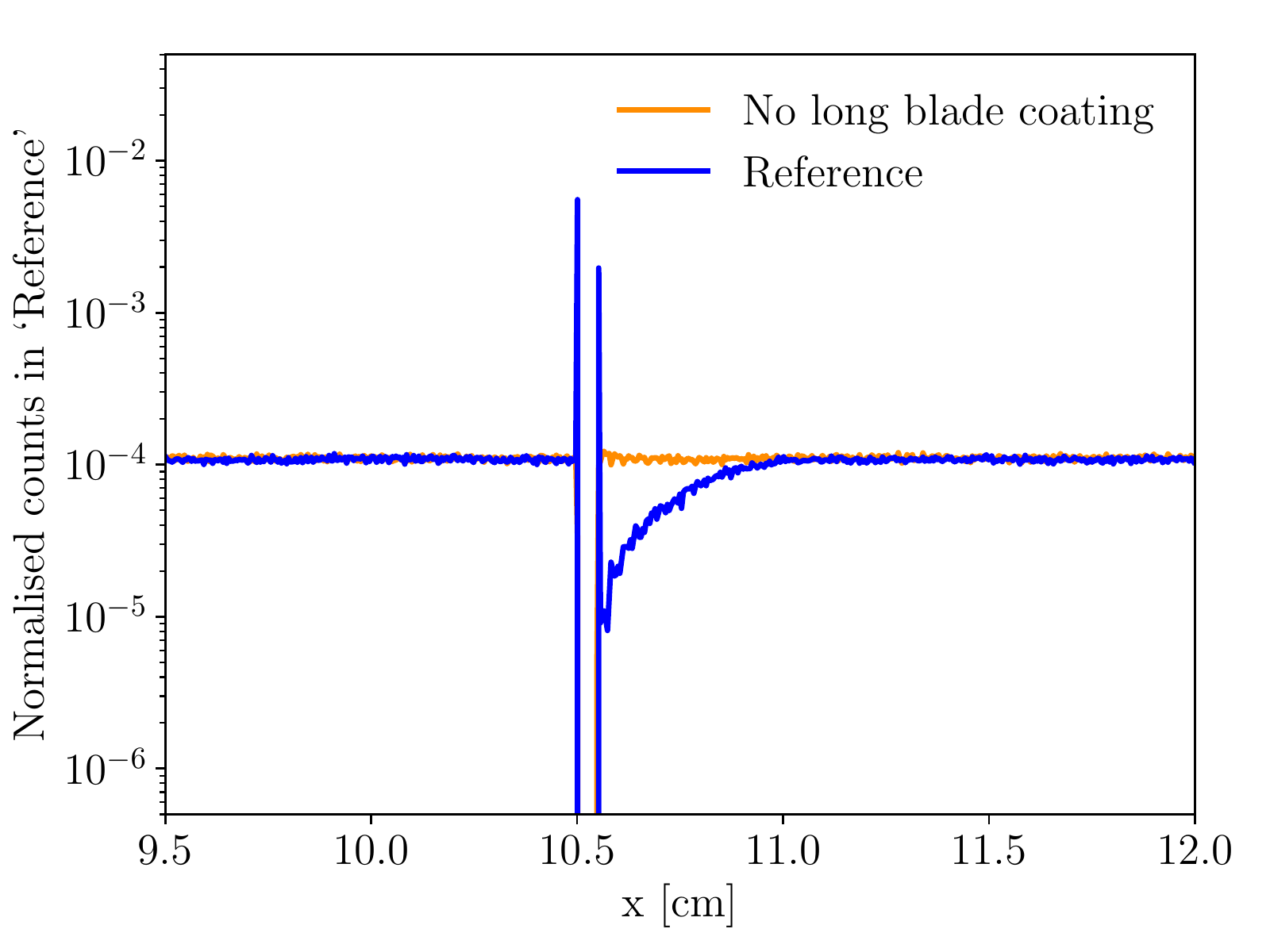}
    \caption{\label{hitxZoom}}
  \end{subfigure}
  \caption{\footnotesize Position of detected neutrons across the width of the detector~(\subref{hitxAll})  and zoomed version~(\subref{hitxZoom}) in the reference detector at 4~\AA~initial neutron wavelength, with and without the long blade coating. 
    Normalised to neutrons entering the detector. The overlapping curves do continue to the rear of the foreground curves. \label{hitx} }  
\end{figure}

\begin{figure}[ht!]
  \centering
  \includegraphics[width=0.75\textwidth]{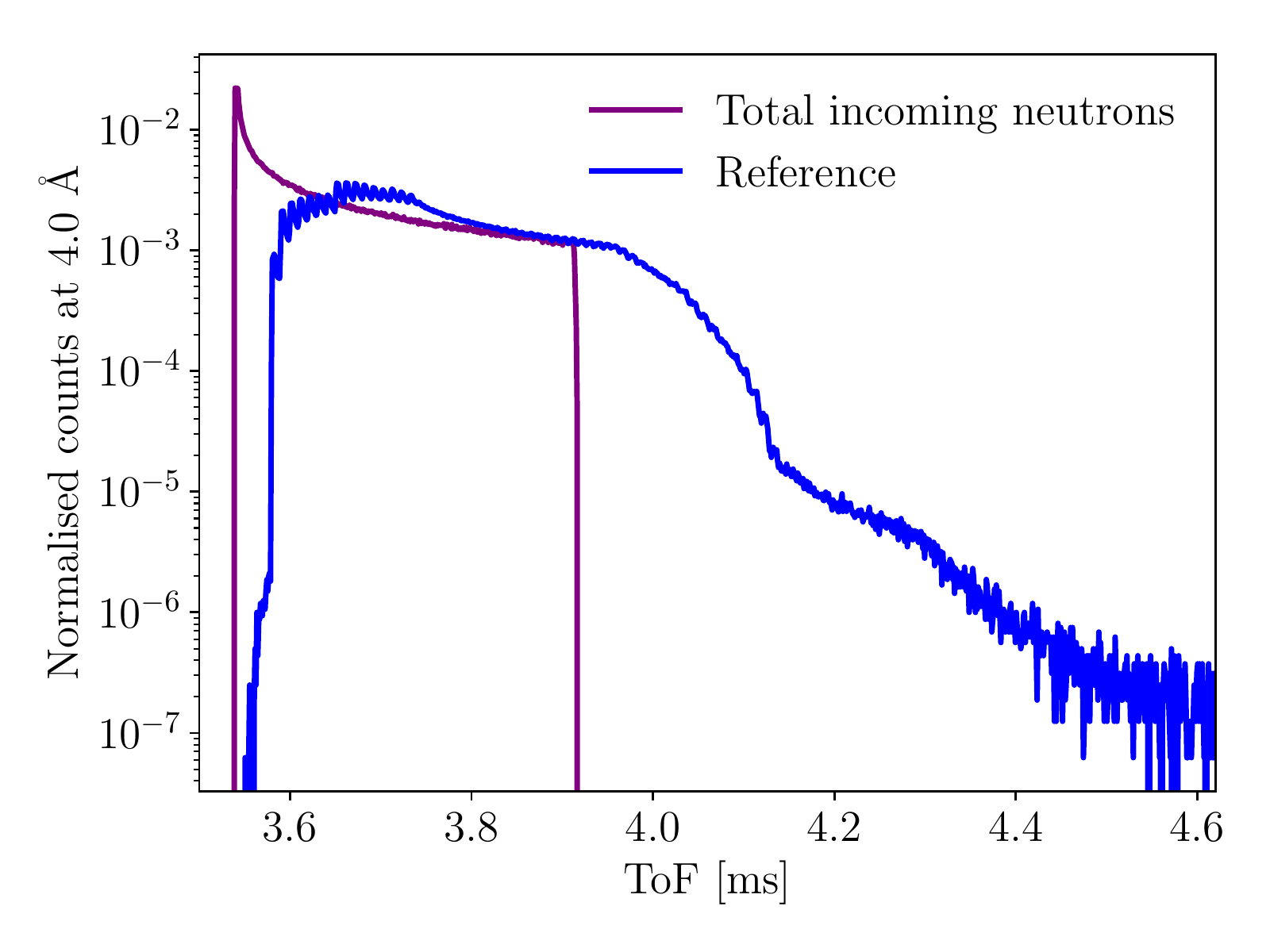}
  \caption{\footnotesize ToF spectrum at 4~\AA~initial neutron wavelength.  Normalised to neutrons entering the detector. The overlapping curves do continue to the rear of the foreground curves. \label{ToF_single}} 
\end{figure}

As the CSPEC is 
an inelastic instrument, the data of interest are the energy- and momentum-transfer, derived from the measured Time-of-Flight (ToF) and the flight distance, calculated in-turn from the detection coordinates. Likewise to real measurements, these parameters are 
accessible in the simulation as well, as it is shown in figure~\ref{hitx} and \ref{ToF_single}. 
In figure~\ref{hitx} the x coordinate of neutron conversion points is presented, as indicated in figure~\ref{g4source}.


This way the implemented grid geometry is clearly reflected in both curves in figure~\ref{hitx}: two separate grids with 6 cells in each, and a 6~mm gap between them. The deep and sharp valleys between the cells are 
attributed to the absorption in the 0.5~mm thick long blades.
The impact of the long blade coating is indicated in the blue curve (in the `reference' detector), especially in figure~\ref{hitxZoom}: detection peaks and valleys appear on the inner and outer side of the long blades, respectively. The cause of this phenomenon is that due to the point source, the neutrons reach the inner side of the long blades in a high incident angle, resulting in a high absorption probability in the inner coating layer (`detection peak'). Because of this, fewer neutrons reach the cell on the opposite side of the respective long blade, leading to an asymmetric shadow, a deficit in neutron conversion in the short blade coating (orange curve) of this neighbouring cell. This way the long blade coating effectively acts as shielding. The maxima of these peaks also change with the x coordinate in the detector, corresponding to the source geometry. 

\FloatBarrier



In figure~\ref{ToF_single} ToF spectra are presented, where `measured' ToF (blue) is defined as the time it takes the neutrons from the sample position to the conversion point in the detector, and the incident, raw ToF (purple) takes from the sample position to entrance window of the detector. The incident ToF spectrum has a sharp start, belonging to the minimum sample-detector distance. The shape of the peak reflects the flight distance distribution from an isotropic point source to the tall, flat window surface, while the width of the spectrum is determined by the longest possible distance, and therefore by the height of the entrance window.  
The conversion ToF spectrum has similar overall characteristics. Here a small background shoulder is present 
before the 3.6~ms edge, containing the neutrons that gained energy in inelastic scattering, appearing with higher velocity in the spectrum. 
The long, falling tail after the peak consists of the elastically scattered neutrons and the ones with energy loss from inelastic scattering, appearing with lower velocity in the spectra. The 
broadening of the ToF peak corresponds again to the height of the detector module, while the tiny peaks, that clearly appear at the beginning, but are smeared over through the whole peak, reflect the parallel conversion layers within the depth of the detector. 




The simulation thus allows access to 
otherwise directly not measurable quantities, e.g. the 
shift of the momentum of the neutrons.

\begin{figure}[ht!]
  \centering
    \includegraphics[width=0.75\textwidth]{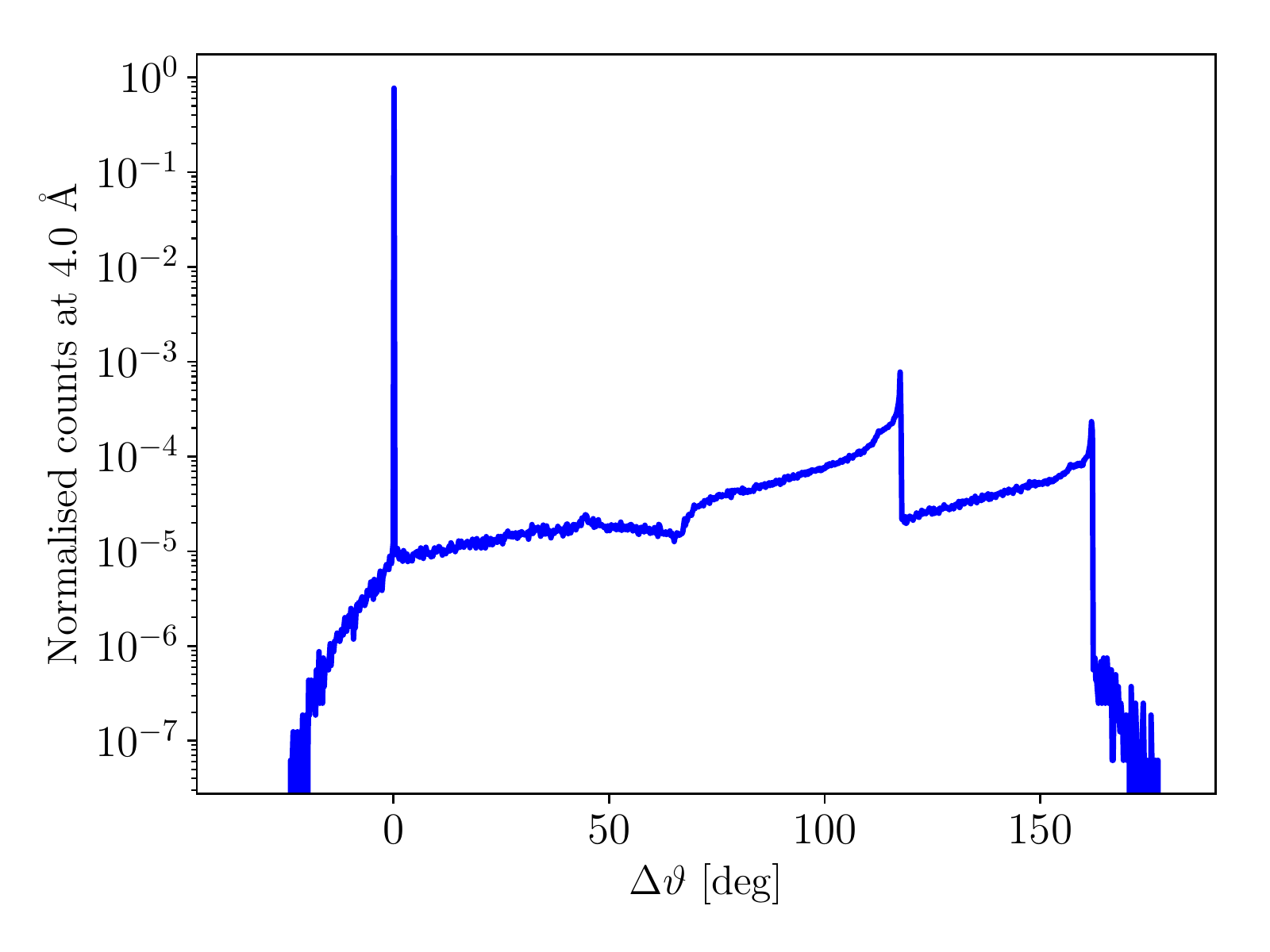}
  \caption{\footnotesize $\Delta \vartheta$ of initial and final polar angle of momentum vector of converted neutrons with 4~\AA~initial wavelength. Normalised to neutrons entering the detector. \label{Dtheta} }   
\end{figure} 

In figure~\ref{Dtheta} the change of the polar angle ($\vartheta$) is presented for 4~\AA~neutrons, defined as $\Delta \vartheta = \vartheta_{final} - \vartheta_{initial}$, where $\vartheta_{final}$ is calculated from the momentum vector of the neutron at conversion point.  A 
sharp peak of non-scattered neutrons is visible at $\Delta \vartheta = 0^{\circ}$, and a continuous scattered neutron background from -25$^{\circ}$ to 180$^{\circ}$. Two asymmetric peaks are also present, 
as at this wavelength, on aluminium, Bragg diffraction can only happen in one of two distinct Debye-Scherrer cones, corresponding to scatter angles of 117$^{\circ}$ and 162$^{\circ}$ respectively.  
This indicates that polycrystalline effects are 
correctly reproduced in the simulation. The peaks are asymmetric due to the 
varying incident angle along the height of the detector.


On the basis of figure~\ref{Dtheta} a precise discrimination of scattered and non-scattered neutrons is feasible: neutrons are taken as non-scattered,  
if  0$^{\circ} \leq \Delta \vartheta \leq 0.4^{\circ}$, which corresponds to the maximum resolution of the detector, 
determined by the subtended solid angle to a cell in the centre front row of the detector. This discrimination allows to define the SBR with only the above defined non-scattered neutrons as signal~(S), while the background~(B) only involves the intrinsic neutron scattering in the detector. In the current study all these quantities are determined for converted neutrons.

\begin{equation}\label{eq:SBR} 
  SBR_{converted \ neutrons} = \frac{ N_{non-scattered}\biggr\rvert_{0^{\circ} \  \leq  \ \Delta \vartheta \ \leq \ 0.4^{\circ}} }{N_{scattered}}  \   \\
\end{equation}

The SBR as defined above is used as a `figure of merit' for the detector design optimisation through this study. It has to be emphasized, that this definition is not the peak to background ratio that can be read from a measured spectrum, but it is calculated from the counts in the area under the peak and the background, respectively. 

\begin{figure}[ht!]
  \centering
  \includegraphics[width=0.75\textwidth]{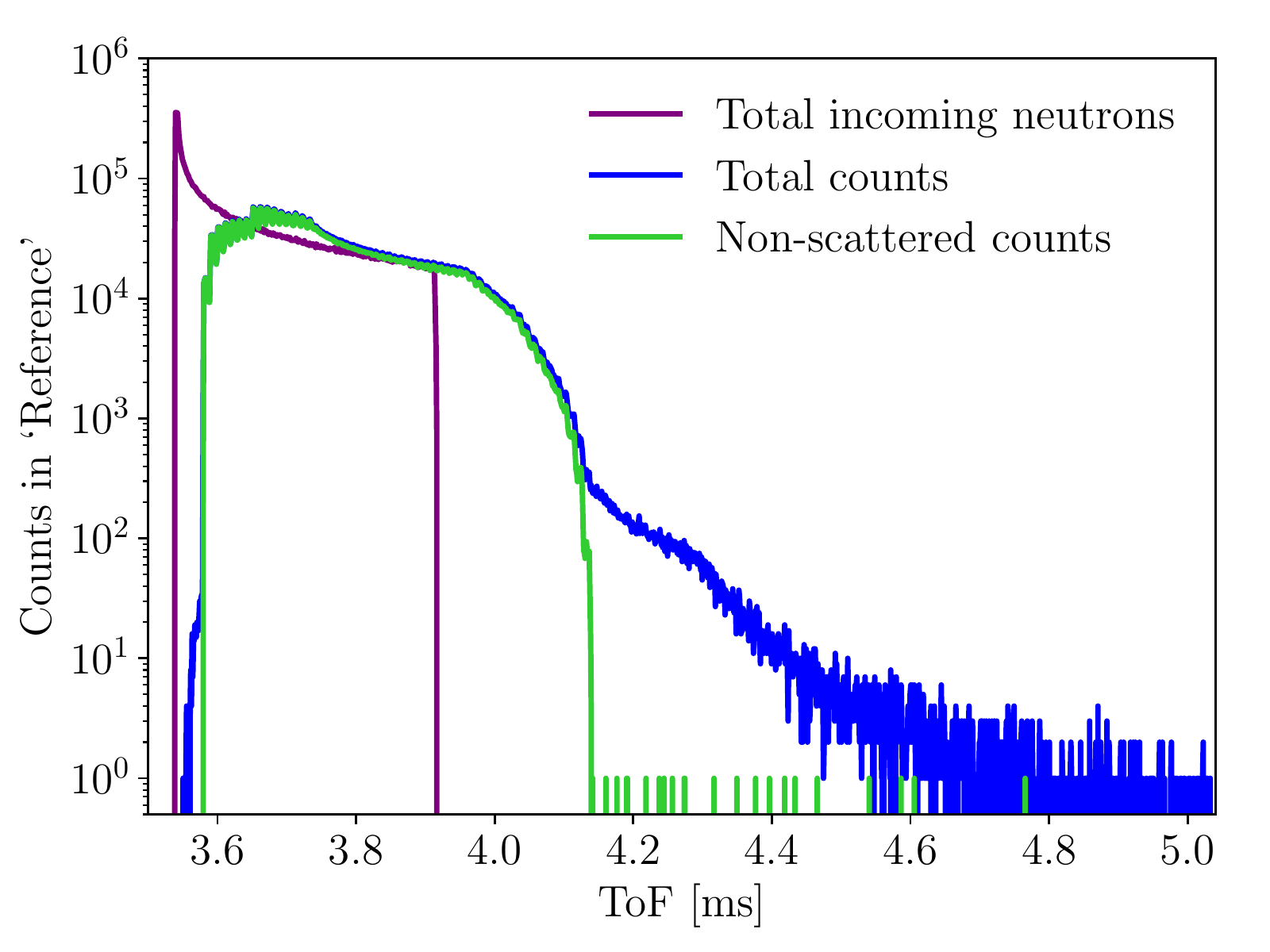}
  \caption{\footnotesize Comparison of ToF spectra from  all and non-scattered neutrons at 4~\AA~initial neutron wavelength.  The overlapping curves do continue to the rear of the foreground curves. \label{scToF}} 
\end{figure}

\begin{figure}[ht!]
  \centering
  \begin{subfigure}[b]{0.49\textwidth}
    \includegraphics[width=\textwidth]{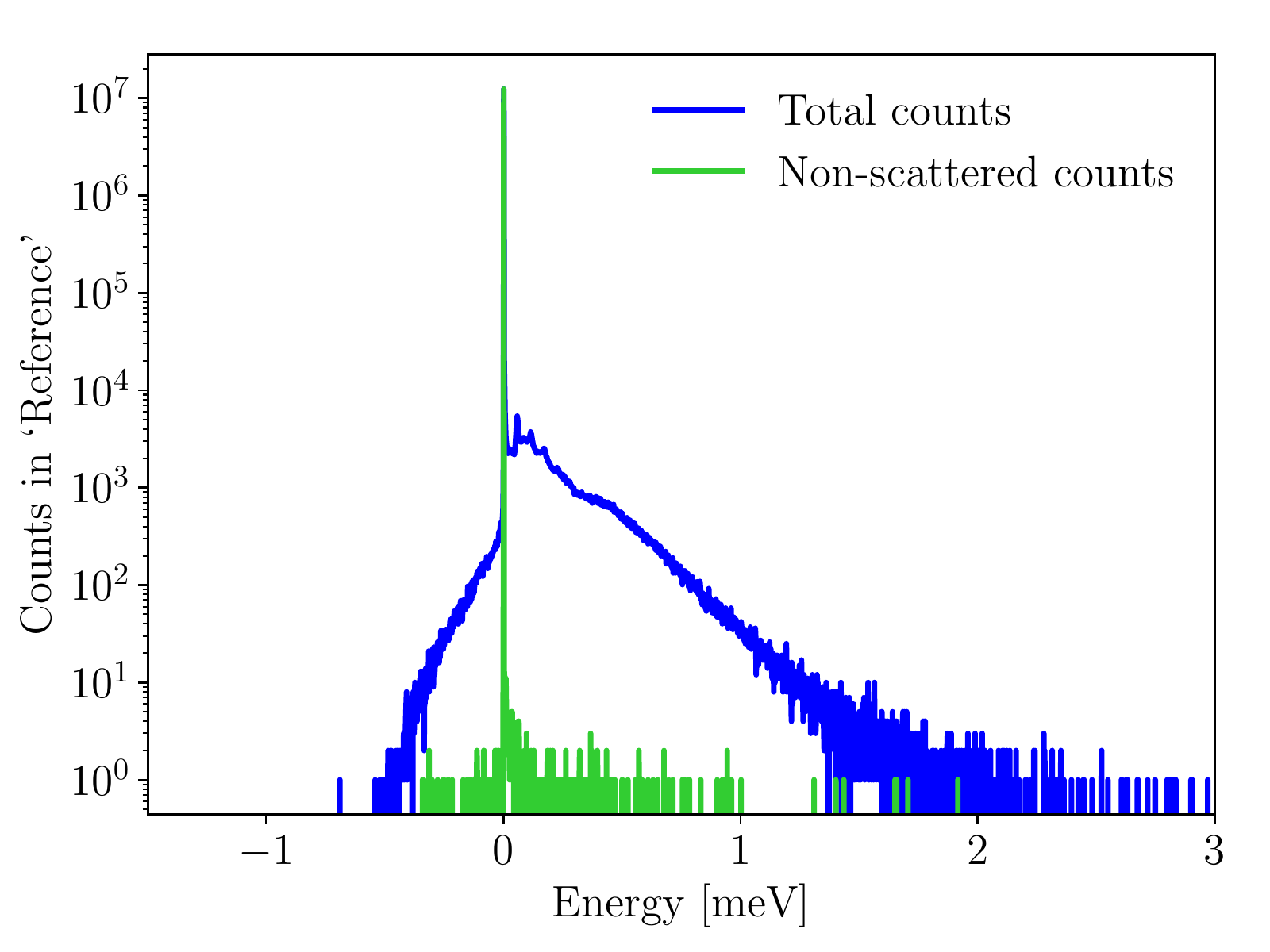}
    \caption{\label{scEtrf_m}}
  \end{subfigure}
  \begin{subfigure}[b]{0.49\textwidth}
    \includegraphics[width=\textwidth]{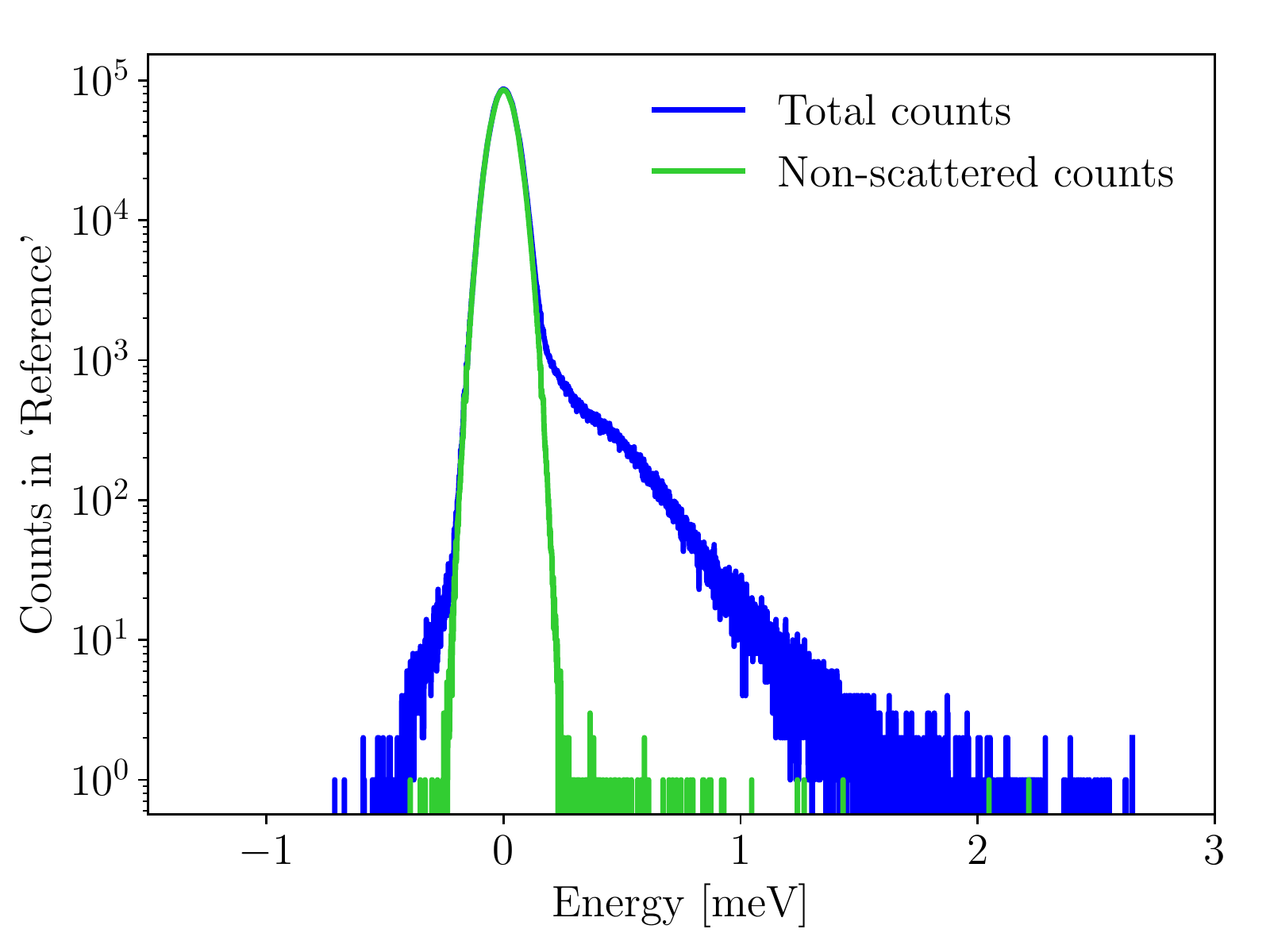}
    \caption{\label{scEtrf_g}}
  \end{subfigure}
 
  \caption{\footnotesize Comparison of energy transfer spectra from all and non-scattered neutrons  with mono-energetic~(\subref{scEtrf_m}) and Gaussian~(\subref{scEtrf_g}) initial neutron energy distribution at 4~\AA. The overlapping curves do continue to the rear of the foreground curves. \label{sccontrol} }  
\end{figure}

In figures~\ref{scToF} and \ref{sccontrol} the comparison of the total and non-scattered ToF and energy transfer spectra are given for 4~\AA~ neutrons, respectively.

The afore-mentioned background 
contributions in the ToF spectrum in figure~\ref{scToF} are clearly identified by this definition, given in equation~\ref{eq:SBR}. The background tails are sharply distinguished, although a continuous flat background is still present in the non-scattered spectrum due to the finite angular range in the signal definition. In figure~\ref{sccontrol} the simulated energy transfer spectra, defined as $E_{trf} = E_{initial} - E_{final}$, following the experimental approach, are 
produced with mono-energetic incident neutrons (figure~\ref{scEtrf_m}) and 
with a 
Gaussian initial energy distribution (figure~\ref{scEtrf_g}) with 1\% standard deviation, typical for these instruments.

 Similarly to the ToF spectrum, in both spectra a smaller fraction of inelastically scattered neutrons appear on the negative side of the spectra, consists of the neutrons that gained energy in scattering, while the shoulder on the positive side consisting of the neutrons that lost energy in inelastic scattering, or had an increased ToF due to elastic scattering, and therefore appear as slower. In the case of the mono-energetic neutrons, minor peaks also appear in the close proximity of the elastic peak on the positive side, belonging to a few rows of backscattering from the short blades within the grid. These peaks are smeared out for the longer 
 flight paths, deeper in the grid structure. In the non-scattered spectra, the signal is again clearly identified, with the presence of the same small, continuous background caused by the finite resolution, that appeared in the ToF spectrum in figure~\ref{scToF}. 
Thus, figure~\ref{sccontrol} confirms the effectiveness of the applied background definition.

%
%
It has to be mentioned that in the more realistic case of the Gaussian initial energy distribution, this discrimination would underestimate the simulated peak, since the scattered neutrons with small $\Delta$E or $\Delta \vartheta$, that still appear within the width of the Gaussian elastic peak would be defined as signal counts in a measurement.



\begin{figure}[ht!]
  \centering
    \includegraphics[width=0.75\textwidth]{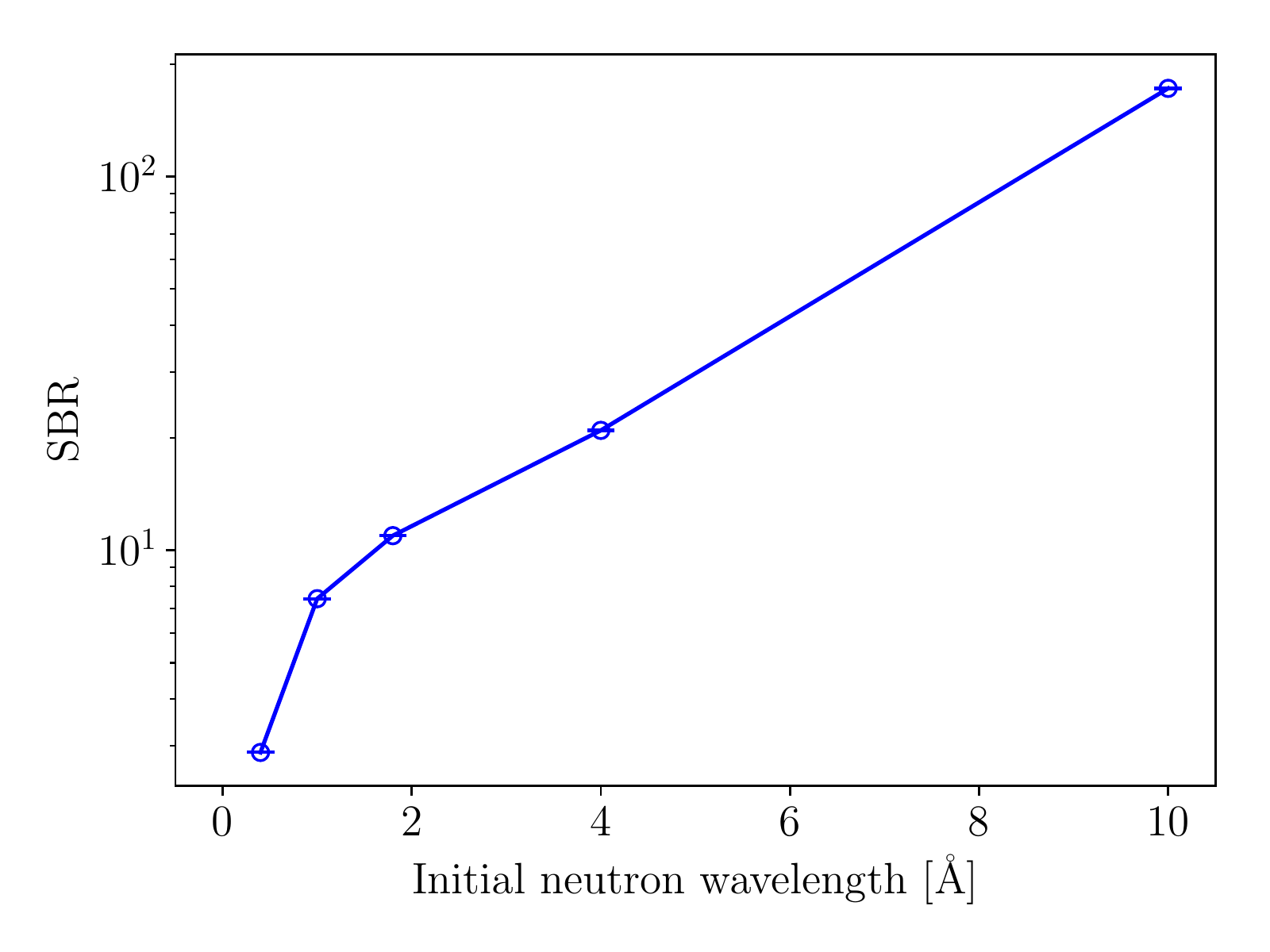}
  \caption{\footnotesize Simulated Signal-to-Background Ratio (as specifically defined in equation~\ref{eq:SBR}) in the unshielded reference detector.  The statistical uncertainties are too small to be discernible.  \label{refSBR} } 
\end{figure}

The impact on the scattered neutron background for all studied components and shielding is studied via the increase of SBR~(SBR$_{rel}$), compared to the SBR of the unshielded reference detector~(SBR$_{ref}$), as it is demonstrated in figure~\ref{refSBR}. 
In this and later upcoming figures, the results of different wavelengths are only connected for better visibility. 

The uncertainties of the simulations are determined and propagated through all the calculations. The simulated signal~(S) and background~(B), as defined for their use in equation~\ref{eq:SBR} are 
independent quantities with Poisson error, and their uncertainties are propagated to 
SBR~(equations~\ref{eq:SBR} and \ref{eq:gerp_sbr}) and relative SBR~(equations~\ref{eq:SBR_rel} and \ref{eq:gerp_rel}) via the Gaussian Error Propagation Law:

\begin{equation}\label{eq:gerp_sbr} 
  \sigma_{SBR} = \sqrt{ \left( \frac{1}{B} \right)^2 \sigma_{S}^2 + \left( \frac{-S}{B^2} \right)^2 \sigma_{B}^2 }\\
\end{equation}

and

\begin{equation}\label{eq:SBR_rel} 
  SBR_{Rel} = \frac{SBR_{\phantom{Ref}}}{SBR_{Ref}},  \\
\end{equation}

\begin{equation}\label{eq:gerp_rel} 
  \sigma_{SBR,Rel} = \sqrt{ \left( \frac{1}{SBR_{Ref}} \right)^2 \sigma_{SBR}^2 + \left( \frac{-SBR}{SBR_{Ref}^2} \right)^2 \sigma_{SBR_{Ref}}^2 }.\\
\end{equation}

Figure~\ref{refSBR} reveals, that the SBR 
monotonously increases with the wavelength of the incident neutrons, and covers a large dynamic range in the operational region of the CSPEC instrument. 
These observations indicate that the proper detector shielding 
has greater significance for thermal neutrons than for cold neutrons.


\FloatBarrier

\section{Scattered neutron background in the CSPEC module design}\label{scBg}


\subsection{Impact of long blade coating }\label{coat}

\begin{figure}[ht!]
  \centering
  \begin{subfigure}[b]{0.49\textwidth}
    \includegraphics[width=\textwidth]{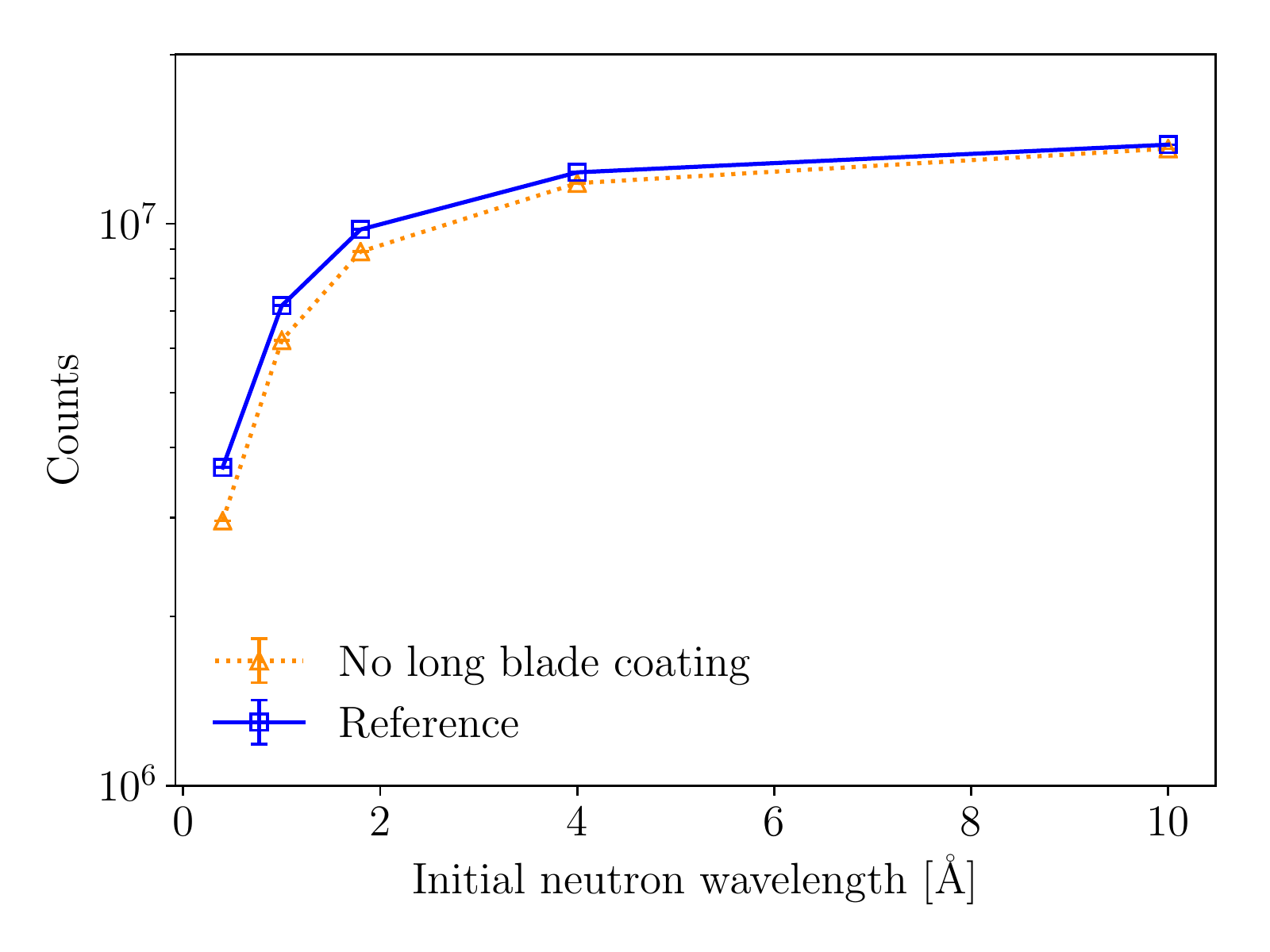}
    \caption{\label{S_lbc}}
  \end{subfigure}
  \begin{subfigure}[b]{0.49\textwidth}
    \includegraphics[width=\textwidth]{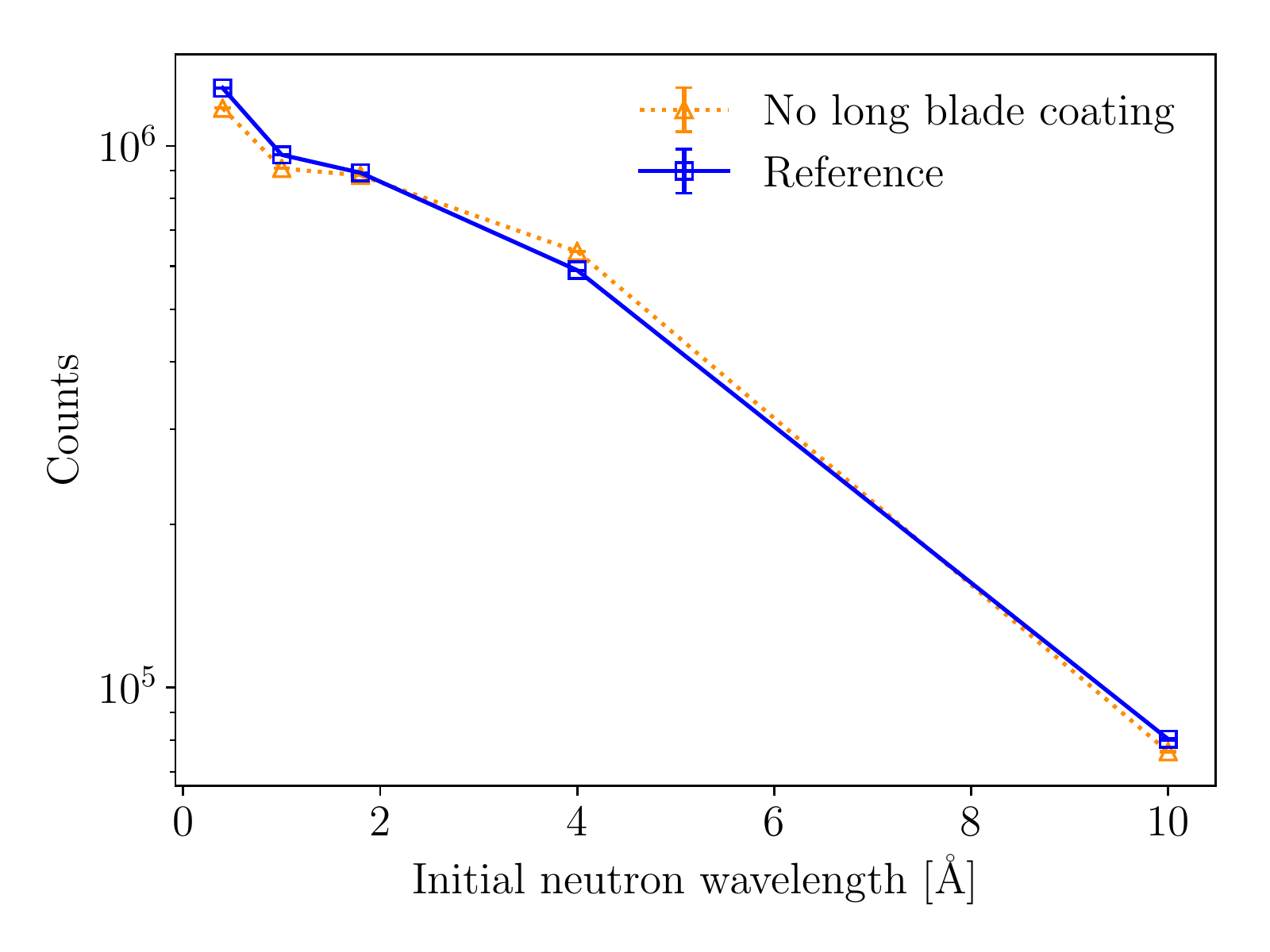}
    \caption{\label{B_lbc}}
  \end{subfigure}
  
  \begin{subfigure}[b]{0.49\textwidth}
    \includegraphics[width=\textwidth]{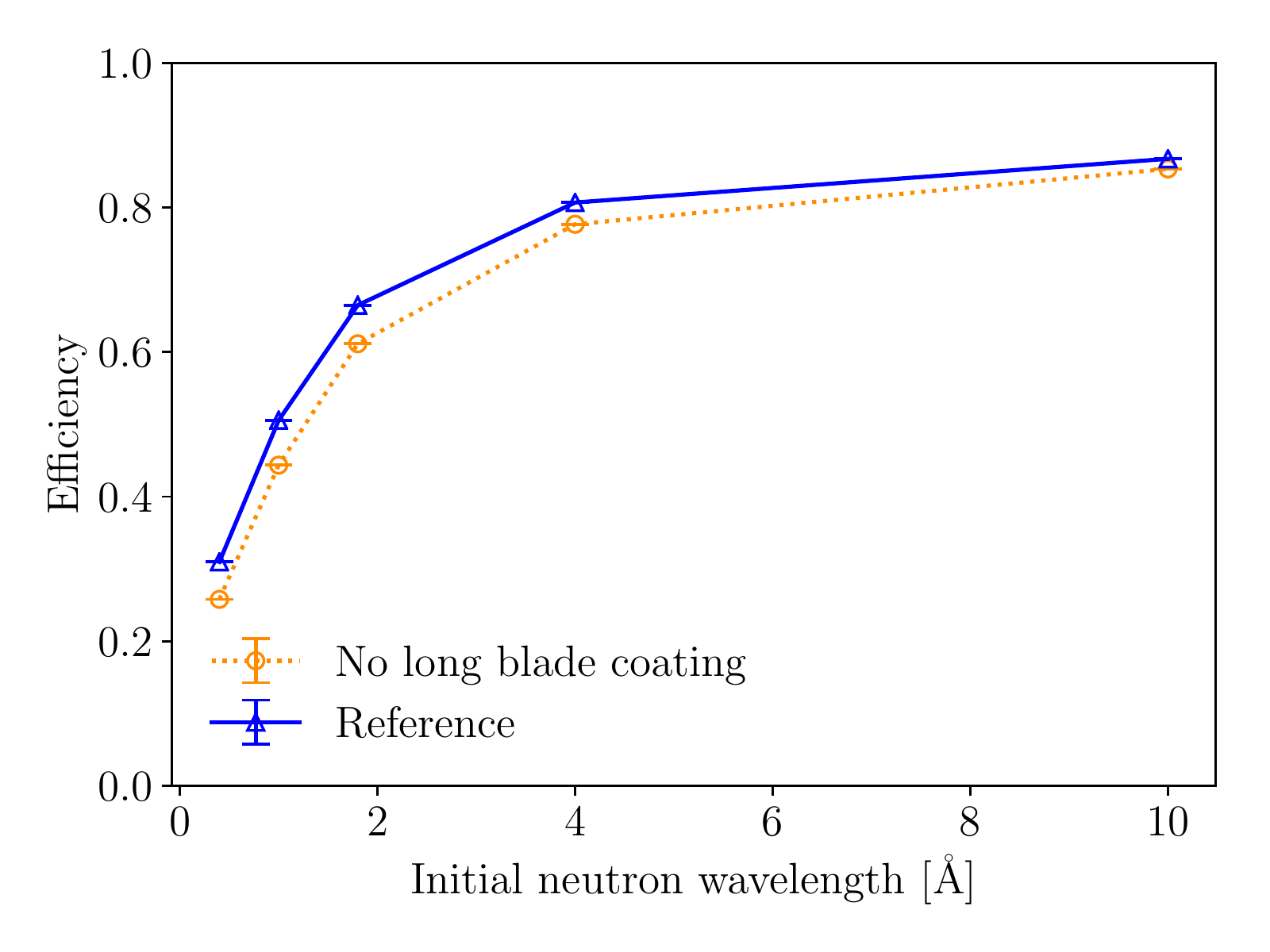}
    \caption{\label{eff_lbc}}
  \end{subfigure}
  \begin{subfigure}[b]{0.49\textwidth}
    \includegraphics[width=\textwidth]{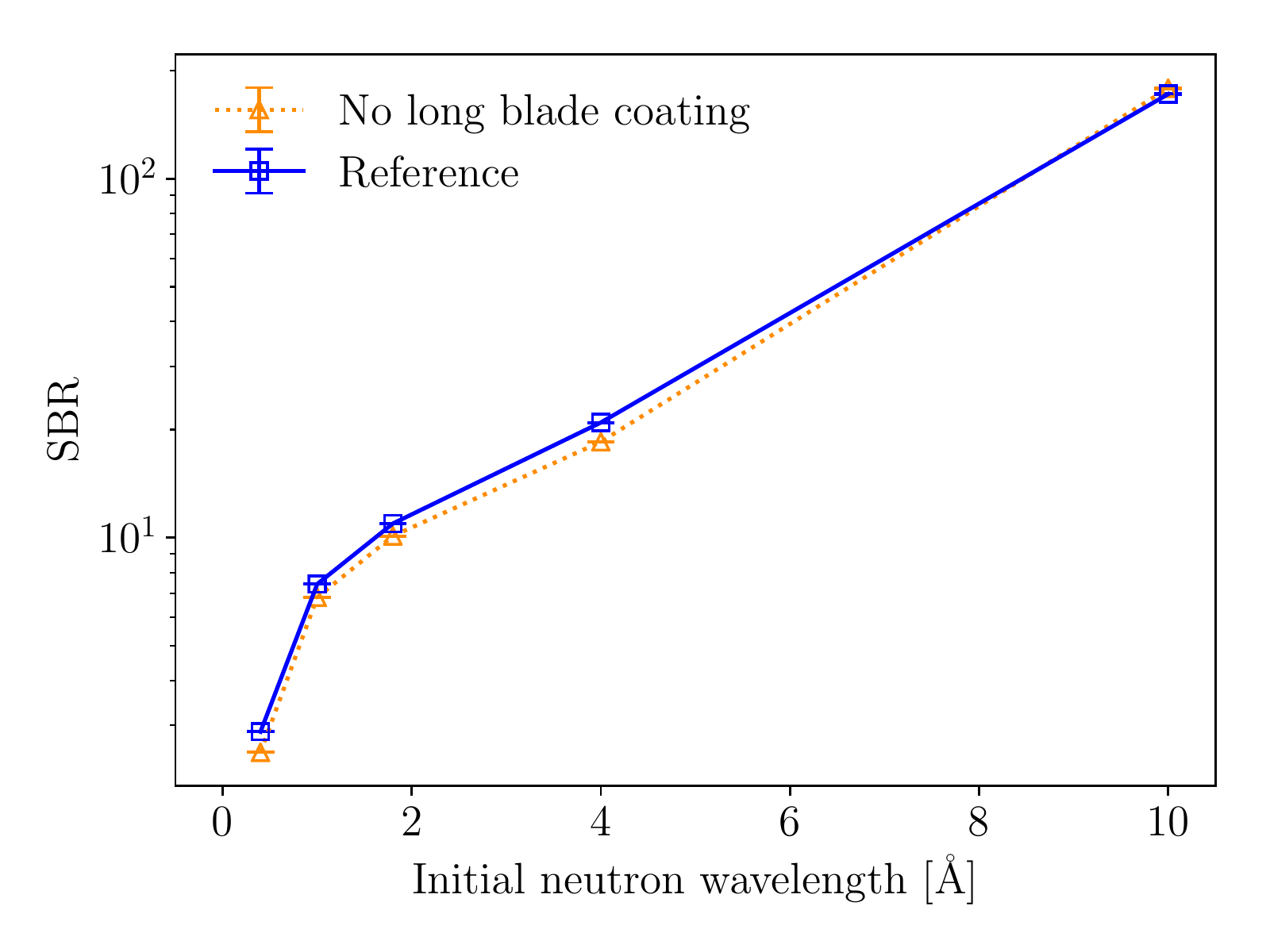}
    \caption{\label{SBR_lbc}}
  \end{subfigure}

  \begin{subfigure}[b]{0.49\textwidth}
    \includegraphics[width=\textwidth]{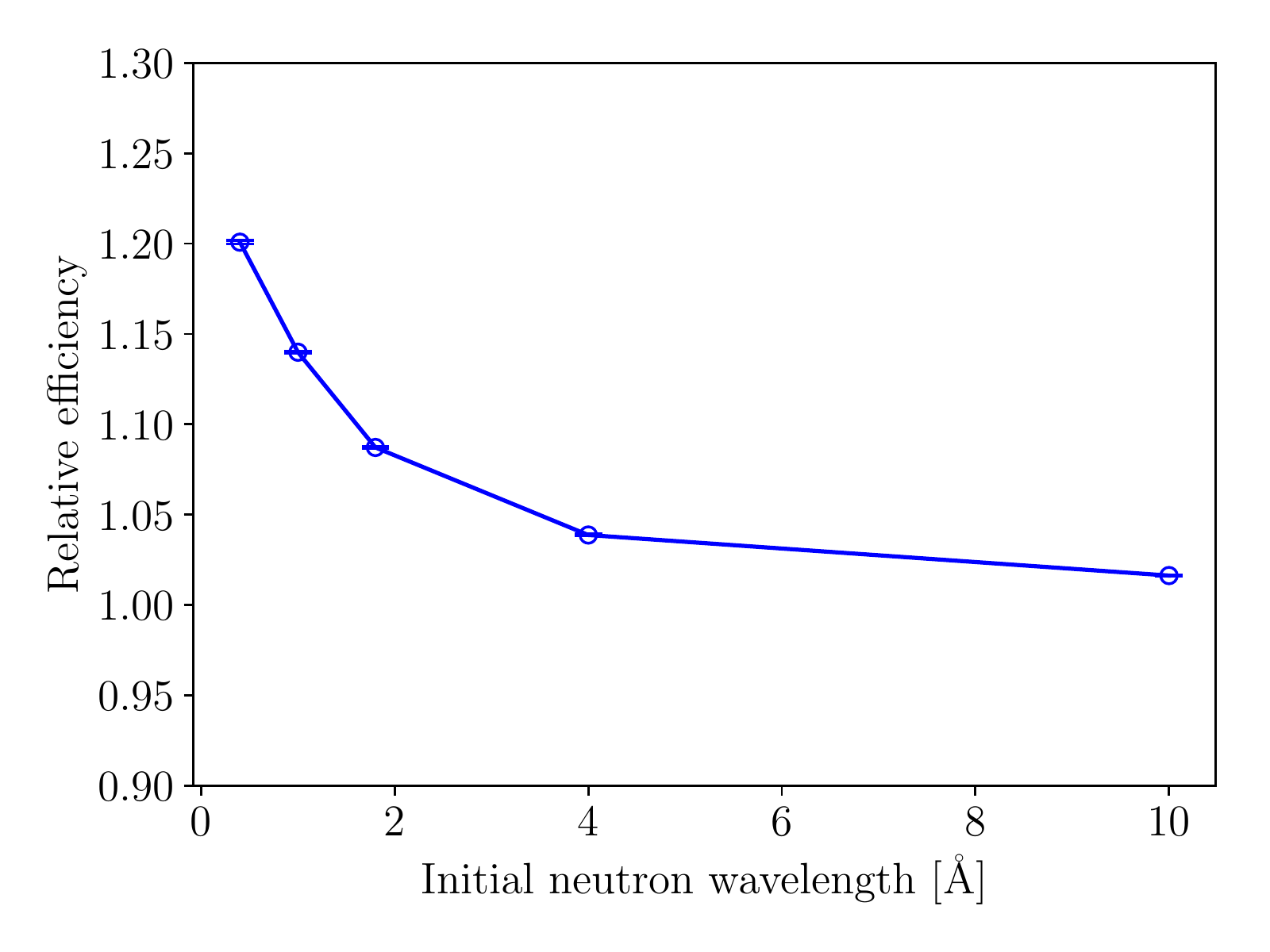}
    \caption{\label{eff_rel_lbc}}
  \end{subfigure}
  \begin{subfigure}[b]{0.49\textwidth}
    \includegraphics[width=\textwidth]{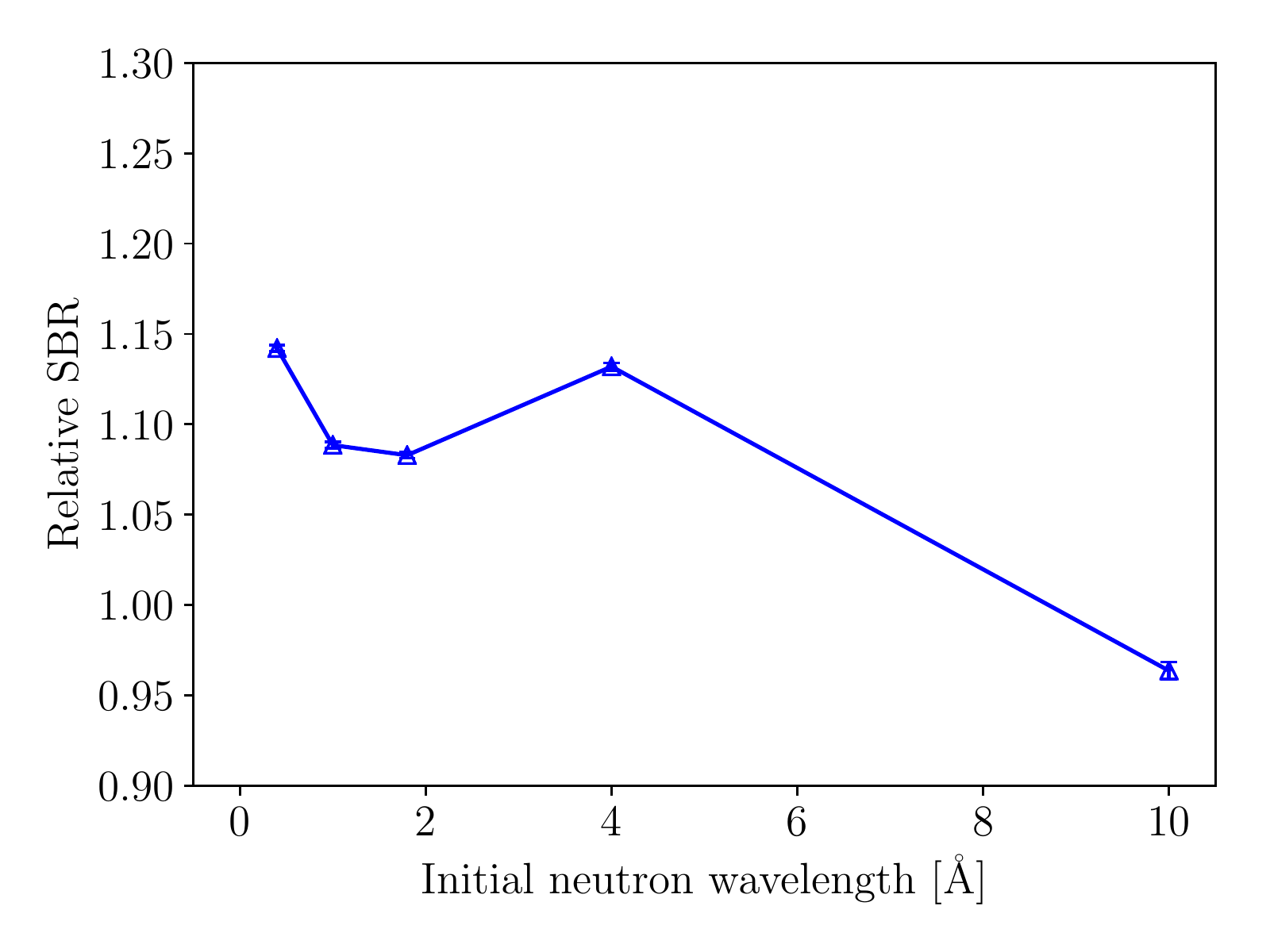}
    \caption{\label{SBR_rel_lbc}}
  \end{subfigure}
  
  \caption{\footnotesize Simulated signal~(\subref{S_lbc}), background~(\subref{B_lbc}), neutron detection efficiency~(\subref{eff_lbc}) and SBR~(as specifically defined in equation~\ref{eq:SBR})~(\subref{SBR_lbc}) as functions of incident neutron wavelength in the unshielded reference detector, with and without coating on the long blades, and change of efficiency~(\subref{eff_rel_lbc}) and SBR~(\subref{SBR_rel_lbc}) with long blade coating compared to no long blade coating case.  The statistical uncertainties are too small to be discernible. \label{lbc} }  
\end{figure}

The planned CSPEC detector module has an improved design in comparison with the thus far built, tested and simulated demonstrators. 
One key difference is that, unlike the previous ones, in this module the long blade of the grids are 
under consideration to be coated with 
1~$\mu$m boron carbide~\cite{brightness_del}. The impact of the long blade coating is studied in the afore-defined reference detector, by comparing the signal~(figure~\ref{S_lbc}), the background~(figure~\ref{B_lbc}), the efficiency~(figures~\ref{eff_lbc}, \ref{eff_rel_lbc}) and the SBR~(figures~\ref{SBR_lbc}, \ref{SBR_rel_lbc}) -- as they are defined in Section~\ref{config} --  
for the wavelengths of interest, with and without the long blade coating. 

Figures~\ref{eff_lbc} and~\ref{eff_rel_lbc} demonstrate a systematic efficiency increase in the presence of long blade coating for all wavelengths, with the increase being more significant, 8-19\%, for the lower conversion efficiencies for the neutrons below 4.0~\AA.
Comparing the SBRs in figures~\ref{SBR_lbc} and~\ref{SBR_rel_lbc}, 
the SBR is increased in the presence of long blade coating at all wavelengths, except in case of 10~\AA, where a negligible 4\% decrease appears in the otherwise highest SBR. The reason of 
this trend is that the long blade coating increases the signal (see figure~\ref{S_lbc}) via the increase of the total efficiency, and affects the scattered neutron background 
via two different aspects of the same process: on the one hand, the background can be increased as the efficiency is increased. 
This effect is dominant at low wavelengths, where the absorption cross section of $^{10}$B is 
low, as it is shown in figure~\ref{B_lbc}.
On the other hand, neutrons more probably get converted in the long blade coating 
before they could scatter e.g.\ on the long blade or on side of the vessel. Therefore, a reduction of background appears at 4~\AA \ in figure~\ref{B_lbc},  where the Bragg-scattering on the aluminium has the dominant impact on the scattered neutron background. This way the shift in the background is determined by the competing reactions of scattering and absorption, and therefore by the respective cross sections of aluminium and $^{10}$B$_{4}$C.

The SBR is determined by the combination of all the impacts on signal and background. In total, 8-14\% increase of SBR can be reached in the 0.4~\AA \ - 4.0~\AA \ wavelength region with the application of 1~$\mu$m long blade coating, as it is presented in figure~\ref{SBR_rel_lbc}.

Due to the positive impact on the efficiency and SBR in the low wavelength region, the application of long blade coating is recommended for the CSPEC instrument, if it can be done at a moderate cost, and sufficient mechanical properties, and should be considered for any instrument 
with respect to the costs and other requirements.

\FloatBarrier

\subsection{Neutron scattering on detector vessel and window}\label{win}

\begin{figure}[ht!]
  \centering
    \includegraphics[width=0.75\textwidth]{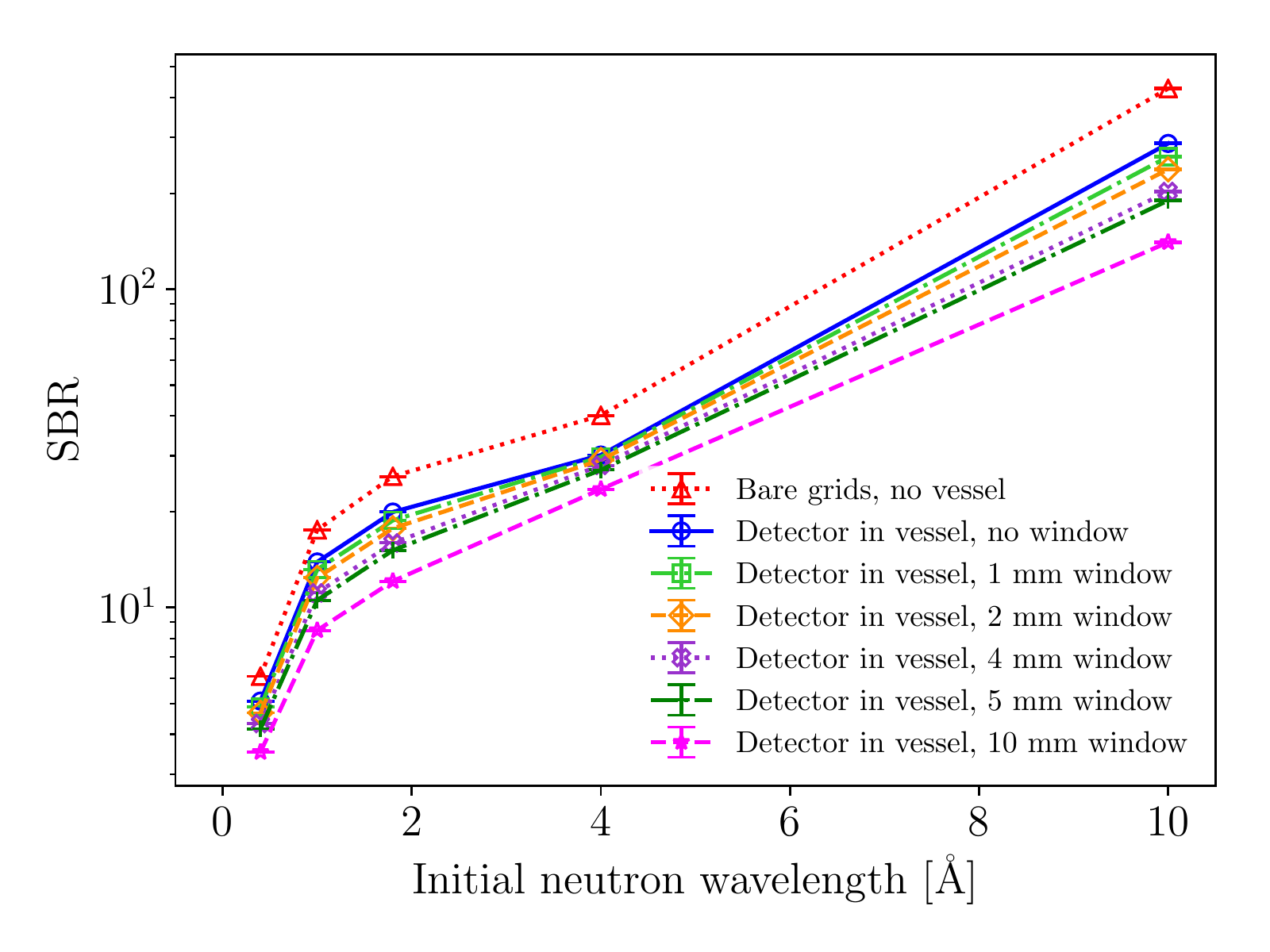}
    \caption{\footnotesize Simulated Signal-to-Background Ratio (as specifically defined in equation~\ref{eq:SBR}) as a function of incident neutron wavelength, with different window thicknesses and vessel components in the reference detector. 
      The statistical uncertainties are too small to be discernible.  \label{winsSBR} }  
\end{figure}

\begin{figure}[ht!]
  \centering
    \includegraphics[width=0.75\textwidth]{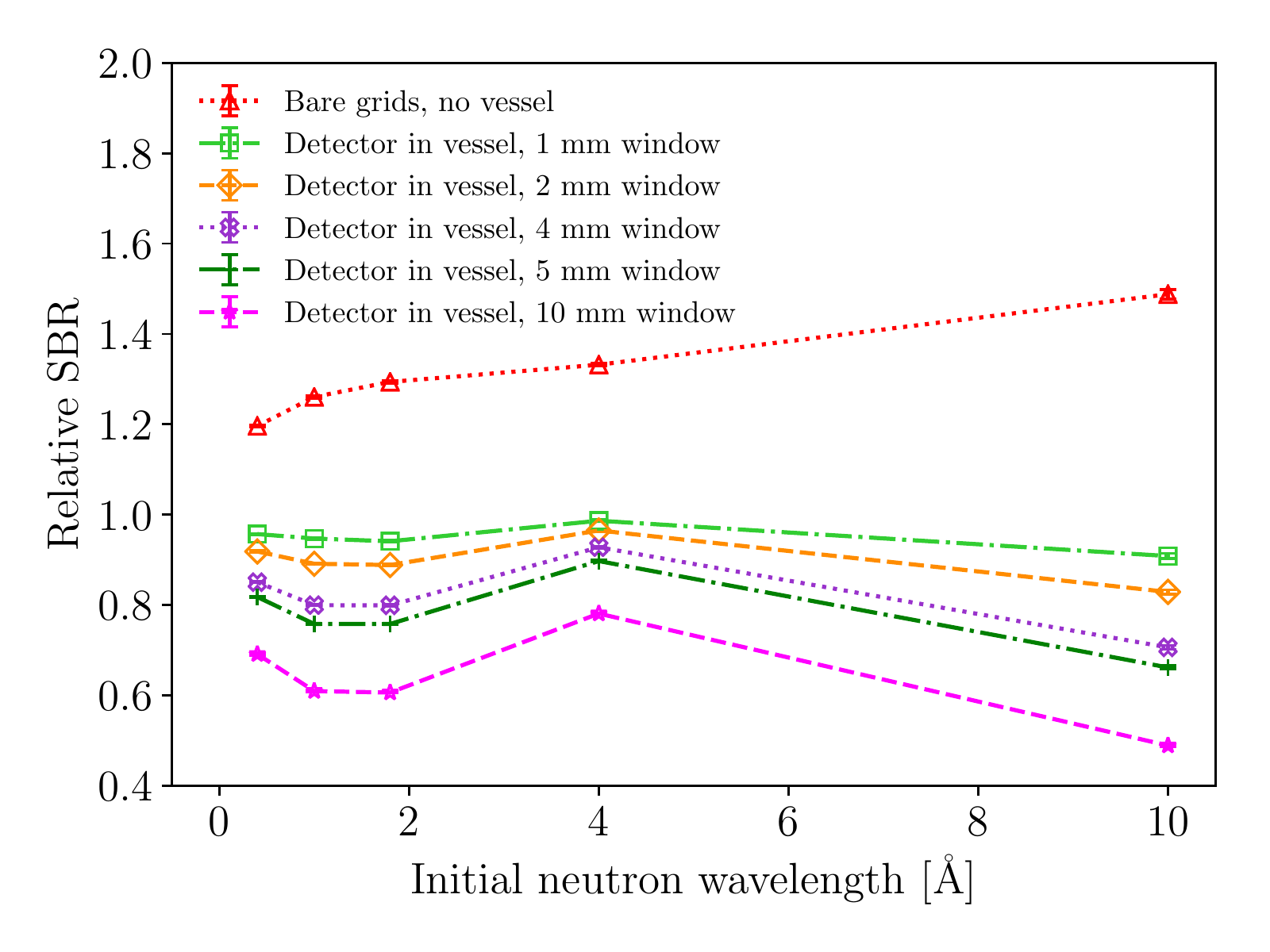}
    \caption{\footnotesize Simulated Signal-to-Background Ratio (as specifically defined in equation~\ref{eq:SBR}) as a function of incident neutron wavelength, with different window thicknesses and vessel components, normalised to the unshielded reference detector 
      with no window.  The statistical uncertainties are too small to be discernible. \label{winsRelSBR} }   
\end{figure}

The scattering on the detector window -- which is an important mechanical structure item as it 
is part of the vacuum interface -- is a well-known 
challange of neutron detector development. 
In the case of the Multi-Grid detector the comparable importance of the scattering on the aluminium vessel has 
been demonstrated in~\cite{dian2018}. Therefore, as first part of the optimisation, the impact of the vessel and the window on neutron scattering is studied. For this purpose, a set of simulations are performed on different versions of the unshielded `reference detector module': 
in the case of the `bare grids', the aluminium vessel is removed. For the other configurations, the vessel is present, but the thickness of the entry window varies between 0~mm (`no window') to 10~mm. The obtained SBRs are presented in figure~\ref{winsSBR}. 
The characteristics of the SBR are the same for all configurations, 
and a continuous decrease of SBR appears with the increased entry window thickness.  Also, except of the bare grids and the 10~mm window case, which 
are 
unphysical and unrealistic cases respectively, no significant difference appears in the SBR. Therefore, to emphasize the discrepancies between the results of the different configurations, in figure~\ref{winsRelSBR} the SBRs are normalised to the no window configuration, 
so the impact of scattering on the window is compared to the respective ideal case.

In figure~\ref{winsRelSBR} it is shown that for realistic window thicknesses (1-5~mm), the decrease of the SBR, compared to the ideal, no window configuration, is less than 10\% at 4~\AA, the wavelength for which the CSPEC instrument is optimised, and the scattering increase is roughly linear with window thickness. The difference is larger in the region where Bragg scattering is dominant, and at long wavelengths: up to 35\% at the most extreme value. However, the incoming wavelength intensity is much larger for intermediate wavelengths around 4~\AA\ at which the scattering is minimal. Given that the no window configuration is unphysical, and that the scattering increase is relatively small for an increase in window thickness, it means that the currently applied 4~mm window, which has been chosen for structural purposes, is confirmed to be a suitable choice between performance and mechanical design optimisation.


On the other hand, it is also revealed that in the bare grids configuration the SBR is increased by 20-50\% due to the lack of scattering on the vessel. The increase of SBR is higher for higher wavelengths, and being 33\% at the optimal 4~\AA. This confirms that a significant increase of the SBR can be 
achieved with suppression of scattering on  the vessel, so internal shielding should be included on the vessel walls (i.e.\ side-shielding). 


\FloatBarrier

\subsection{Study of scattered neutron background suppression with black shielding}\label{black}

The complex structure of the Multi-Grid detector is proven to be a source of 
intrinsic scattered neutron background~\cite{khaplanov2014,dian2018}. However, this complex geometry also leaves space for background reduction via optimisation of shielding design. To obtain an optimised shielding design, the  background reduction capacity of the potential shielding geometry has to be determined. For this purpose black material is applied for each afore-mentioned shielding geometry, to study their impact on the SBR through the whole 0.4-10.0~\AA, which is operational range of the CSPEC instrument. With the application of the black material, the 
highest obtainable background reduction can be determined for each shielding geometry. 
For this purpose different shielding topologies are applied both individually and in combination in the reference detector.

\begin{figure}[ht!]
  \centering
    \includegraphics[width=0.75\textwidth]{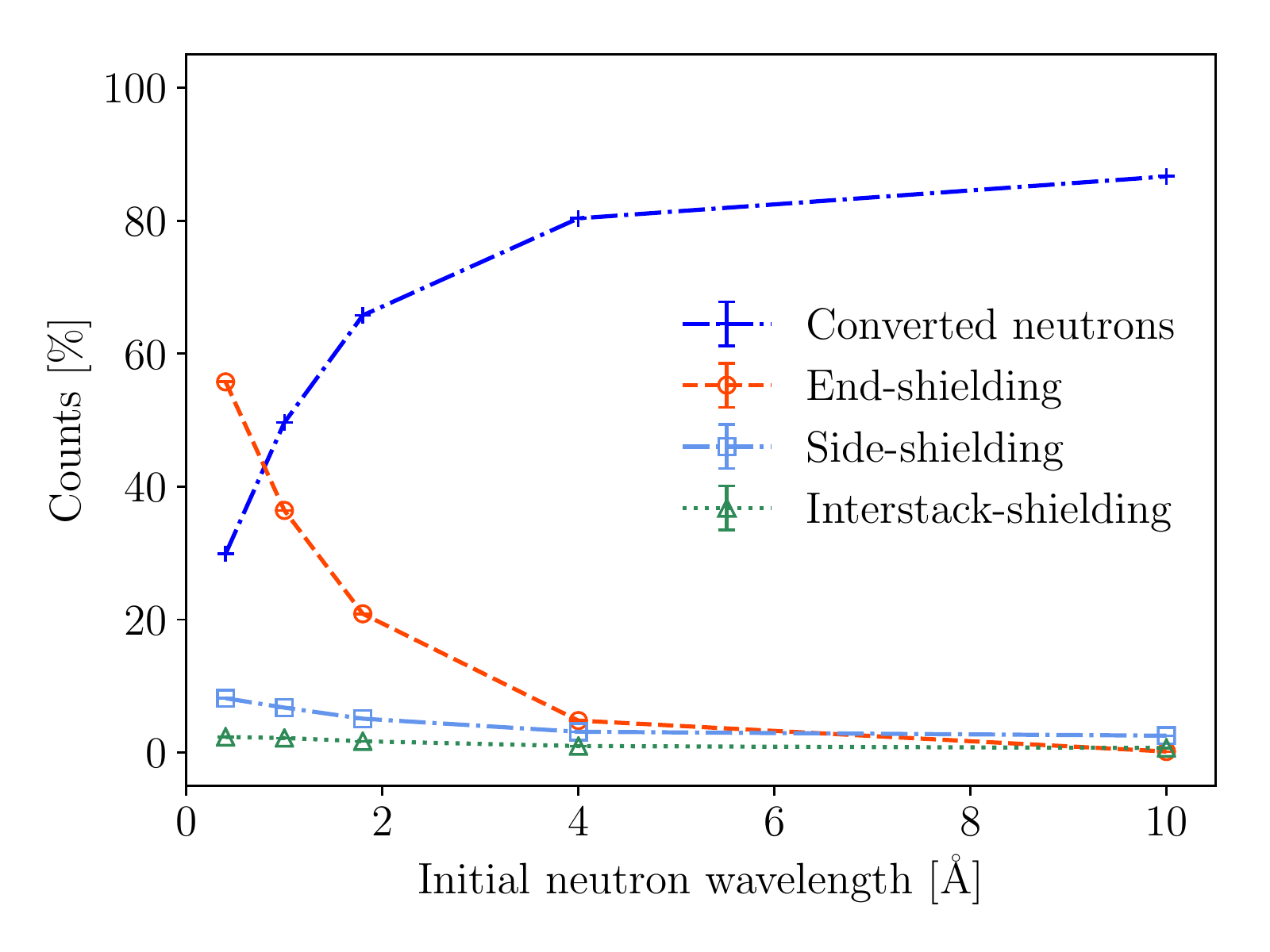}
    \caption{\footnotesize Neutron conversion and absorption in different shielding topologies with black material in the reference detector. 
      The different shielding topologies are applied individually.  The statistical uncertainties are too small to be discernible.  \label{ShCounts}  }    
\end{figure}

The evaluation of the background reduction capacity is performed based on the number of neutrons absorbed in every shielding volume, normalised to the incident neutrons (figure~\ref{ShCounts}). The neutrons absorbed by the converter are also displayed for the sake of completeness. 

The neutron absorption in the end-shielding and the neutron conversion have similar, but 
opposing trends throughout the wavelength range. The reason for this that these are 
competing processes; as the neutron absorption cross-section in the boron carbide converter increases with the wavelength, more neutrons are converted, and 
fewer neutrons reach the end-shielding, so 
fewer neutrons are absorbed in the end-shielding. However, the end-shielding can absorb a significant 
amount of neutrons 
below 4~\AA, and 56\% of the neutrons can be absorbed in the end-shielding at 0.4~\AA. On the other hand at 10~\AA \ the absorption in the end-shielding is practically zero, as the neutrons do not reach the end of the grid.
It is also shown, that the absorption is more even in the other two geometries, i.e.\ 8-2.5\% of the neutrons are absorbed in the side-shielding, 8\% at 0.4~\AA,  and  less than 2\% is absorbed in the interstack-shielding.

Consequently, the end-shielding is the dominant shielding topology in the Multi-Grid detector when black absorber is applied, and there is a high background suppression potential effective in the low-wavelength region, where the SBR is the smallest.





\begin{figure}[ht!]
  \centering
    \includegraphics[width=0.75\textwidth]{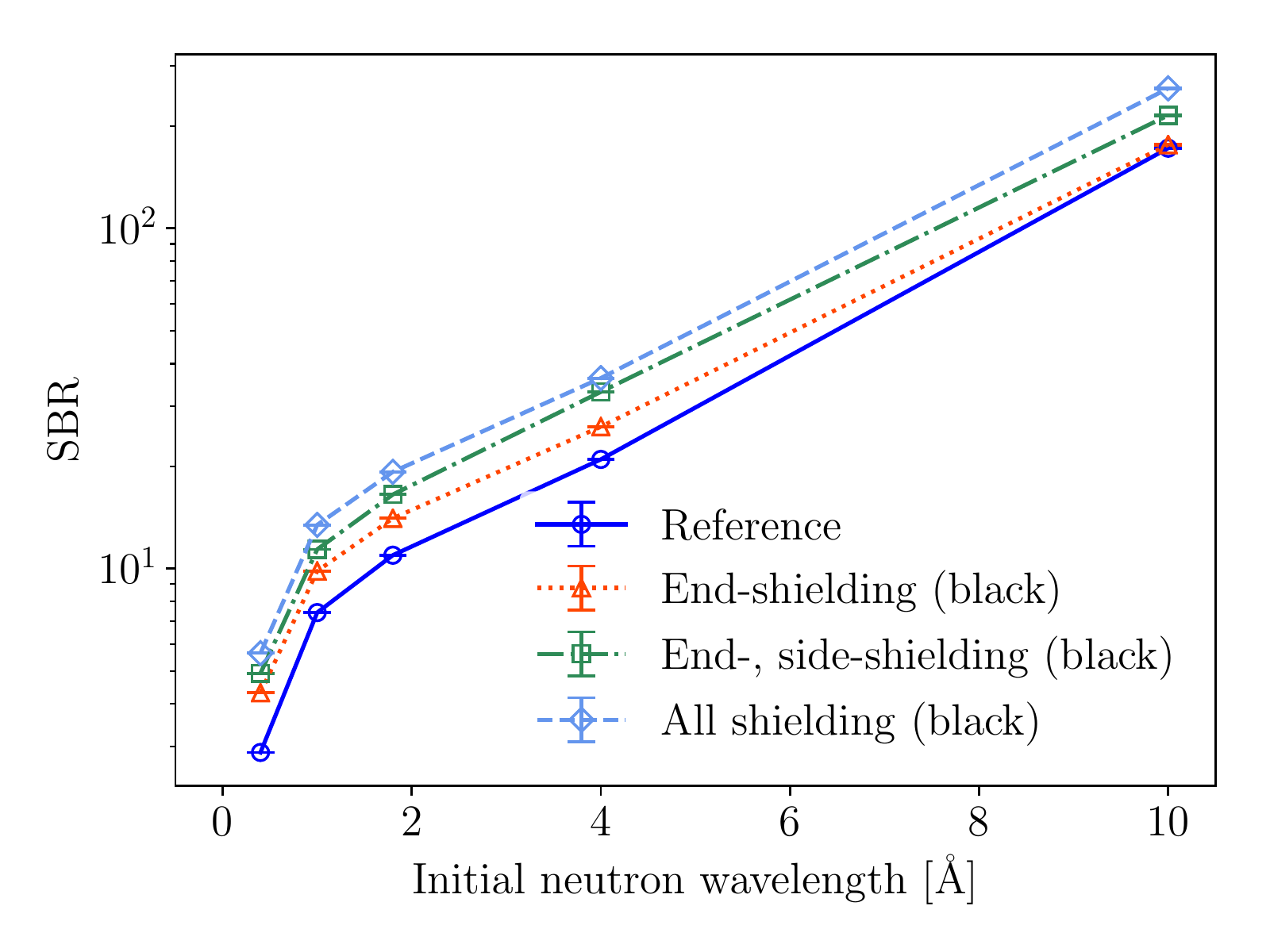}
    \caption{\footnotesize Simulated Signal-to-Background Ratio (as specifically defined in equation~\ref{eq:SBR}) in presence of different shielding topologies with black material in the reference detector. 
      The different shielding topologies are applied in combination.  The statistical uncertainties are too small to be discernible. \label{BlacksSBR}  }  
\end{figure}

\begin{figure}[ht!]
  \centering
  \begin{subfigure}[b]{0.49\textwidth}
    \includegraphics[width=\textwidth]{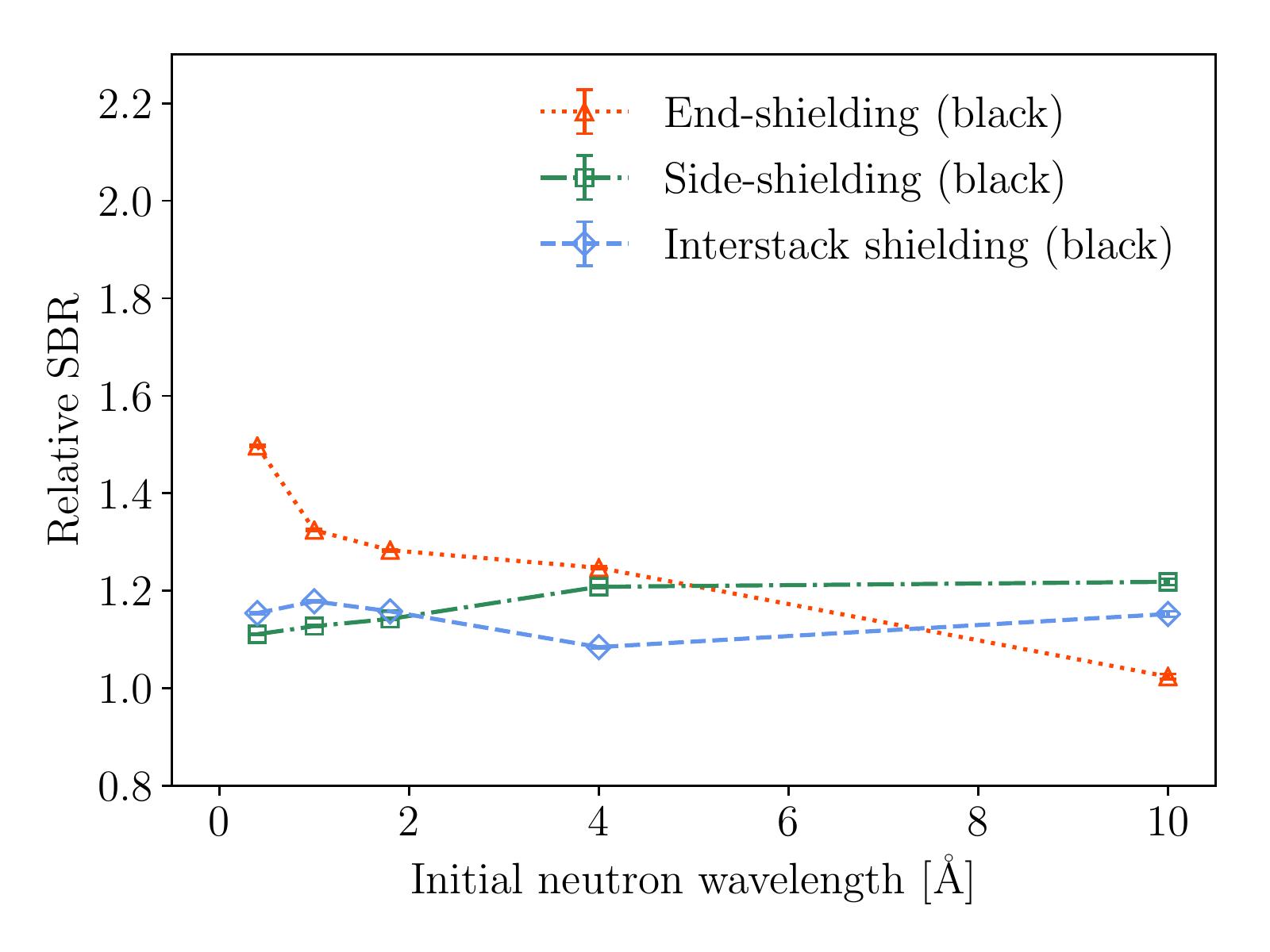}
    \caption{\label{SBR_rel_sBlack}}
  \end{subfigure}
  \begin{subfigure}[b]{0.49\textwidth}
    \includegraphics[width=\textwidth]{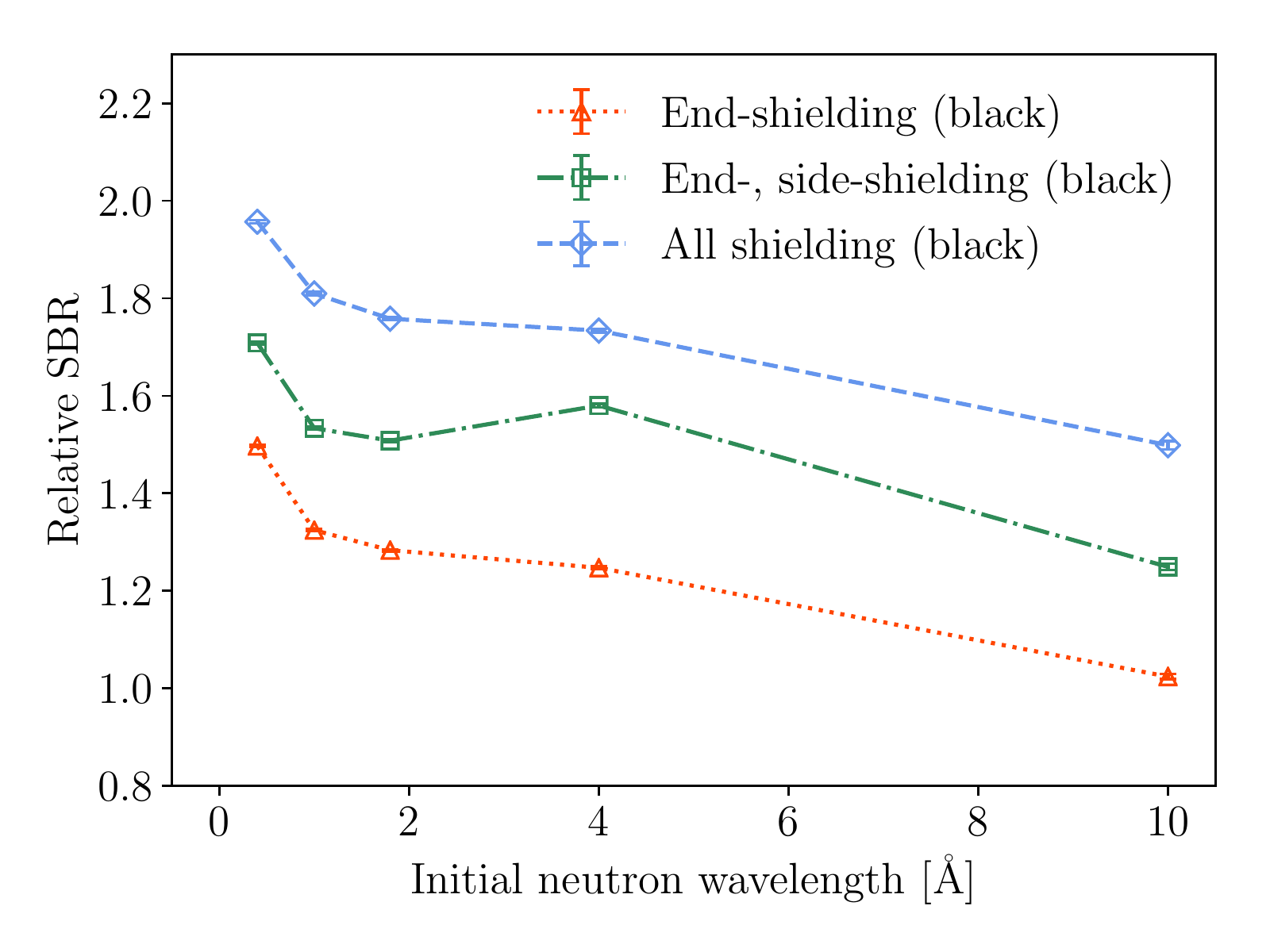}
    \caption{\label{SBR_rel_mBlack}}
  \end{subfigure}

  \caption{\footnotesize Simulated Signal-to-Background Ratio (as specifically defined in equation~\ref{eq:SBR}) in presence of different shielding topologies with black material, normalised to the unshielded reference detector. 
    The different shielding topologies are applied individually~(\subref{SBR_rel_sBlack}) and in combination~(\subref{SBR_rel_mBlack}).  The statistical uncertainties are too small to be discernible.  \label{BlacksRelSBR}  }

\end{figure}

The effect of the different shielding topologies on the SBR is also determined, as it is shown in figure~\ref{BlacksSBR}. The shielding topologies are added one-by-one to the simulation, starting from the reference detector. 
The SBR is increased in the whole wavelength range; the end-shielding has a high contribution in the low wavelength region, while the other two shielding geometries are responsible for the increase of SBR at high wavelengths. The relative increase of the SBR compared to the SBR of the unshielded reference detector 
is depicted in figure~\ref{BlacksRelSBR}, both for individually applied shielding topologies~(\ref{SBR_rel_sBlack}) and for their combinations~(\ref{SBR_rel_mBlack}).

In figure~\ref{SBR_rel_sBlack} the significance of the end-shielding is confirmed; the SBR is increased with 50-25\% in the 0.4-4.0~\AA \ region with the application of black material. Moreover, the increase is the highest in the low wavelength region, where the SBR otherwise is the lowest. The increase of SBR due to the presence of side-shielding is 11-22\%, and 8-18\% due to the interstack-shielding. 
%
It is important to highlight the opposite impact of the side-shielding and the interstack-shielding, that is most significant at 4.0~\AA; the side shielding has a higher impact at this wavelength and above, as the  isotropic scattering becomes the dominant source of background. At 4.0~\AA \ the majority of the scattered neutron background still comes from the Bragg-scattering from the rear blade of the grid. As one of the respective angle of the Bragg-scattering is 
117$^{\circ}$, the scattered neutrons in the Bragg-cone are targeted towards the vessel sides with 
63$^{\circ}$ opening angle, so these neutrons do not reach the interstack-shielding. Therefore the interstack-shielding only absorbs the minority of neutrons scattered in the inner blades of the grids, and has a low impact at  4.0~\AA, but has higher impact at lower wavelength, where the 
overall background is higher due to the lower absorption cross section of $^{10}$B. All these 
phenomena emphasize the potential of a combined shielding design.

In figure~\ref{SBR_rel_mBlack} the different shielding topologies are added one-by-one to the reference detector 
in order of their individually shown relevance. It is shown that the trend of the increase of the SBR is determined by the effect of the end-shielding, although the combination of the end-shielding and the side-shielding has a peak of SBR increase at 4.0~\AA, the optimal wavelength of the CSPEC instrument. This 
effect is caused by the side-shielding, as it is explained in the case of the individual shielding topologies in figure~\ref{SBR_rel_sBlack}. However, 
it is magnified in the combination of shieldings.
In the presence of all three shielding topologies, the trend of the SBR increase is again 
directed by the end-shielding, with a uniform increase along the whole studied wavelength range. It is demonstrated, that the SBR can be increased by up to 25-71\% with the combination of the end-shielding and the side-shielding, by 71\% and 58\% 
at 0.4 and 4.0~\AA, respectively. Also, the SBR can be increased up to 50-96\% with the combination of all three shielding topologies, and by 96\% and 73\% at 0.4 and 4.0~\AA, respectively.
To sum up, complex shielding design is confirmed to have a remarkable potential to increase the SBR via background suppression. This impact is higher in the low wavelength region, where the efficiency and the SBR are inherently lower, and it is proven, that an ideal, combined shielding has the potential to increase the SBR by 50\% at the  4.0~\AA \ operating optimum of the CSPEC instrument, and to almost double it at 0.4~\AA.

\FloatBarrier

\subsection{Shielding optimisation in a Multi-Grid detector module for CSPEC instrument at ESS}\label{sh}

\begin{figure}[t!]
  \centering
  \begin{subfigure}[b]{0.49\textwidth}
    \includegraphics[width=\textwidth]{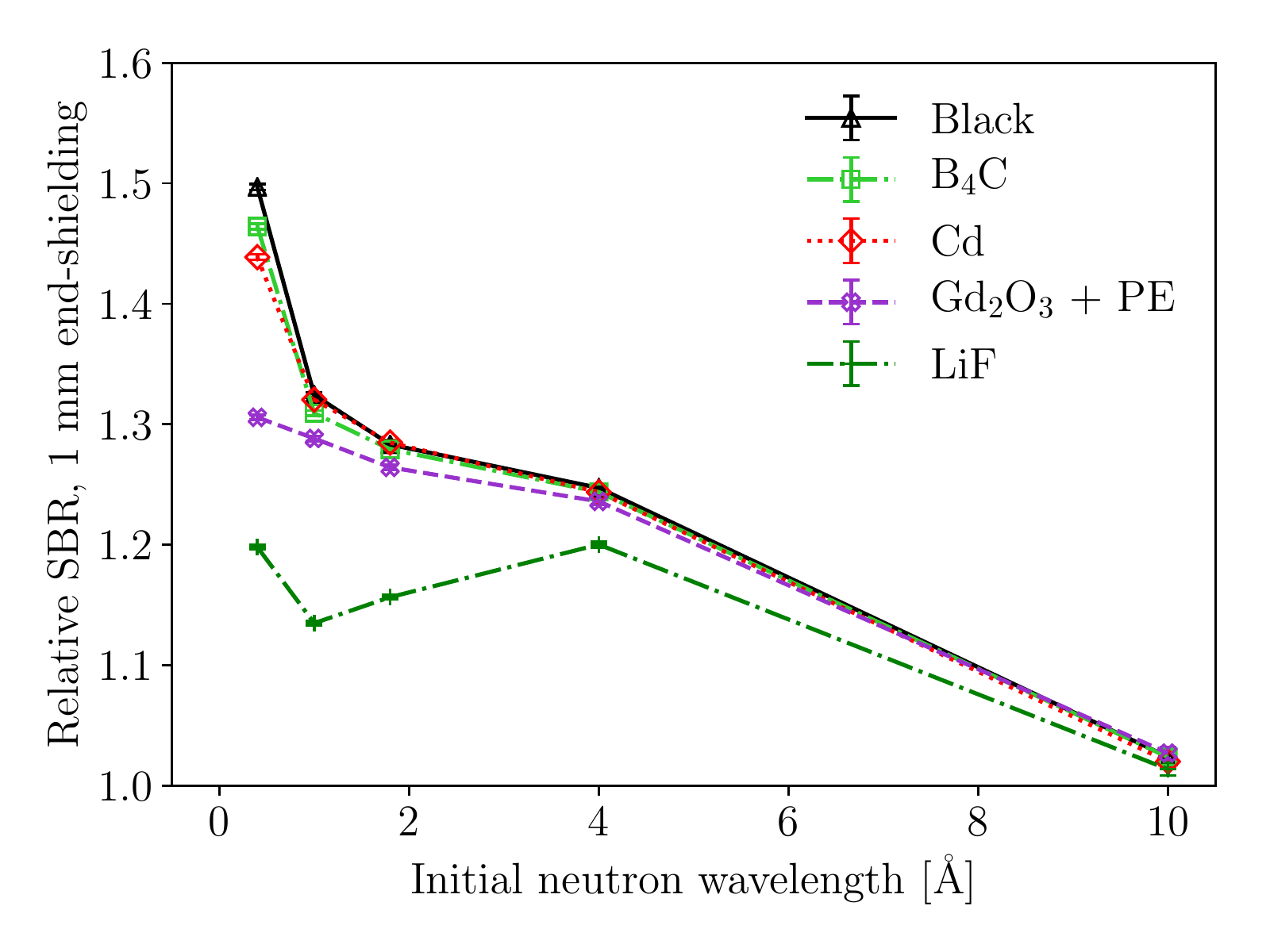}
    \caption{\label{MatGridEnd}}
  \end{subfigure}
  \begin{subfigure}[b]{0.49\textwidth}
    \includegraphics[width=\textwidth]{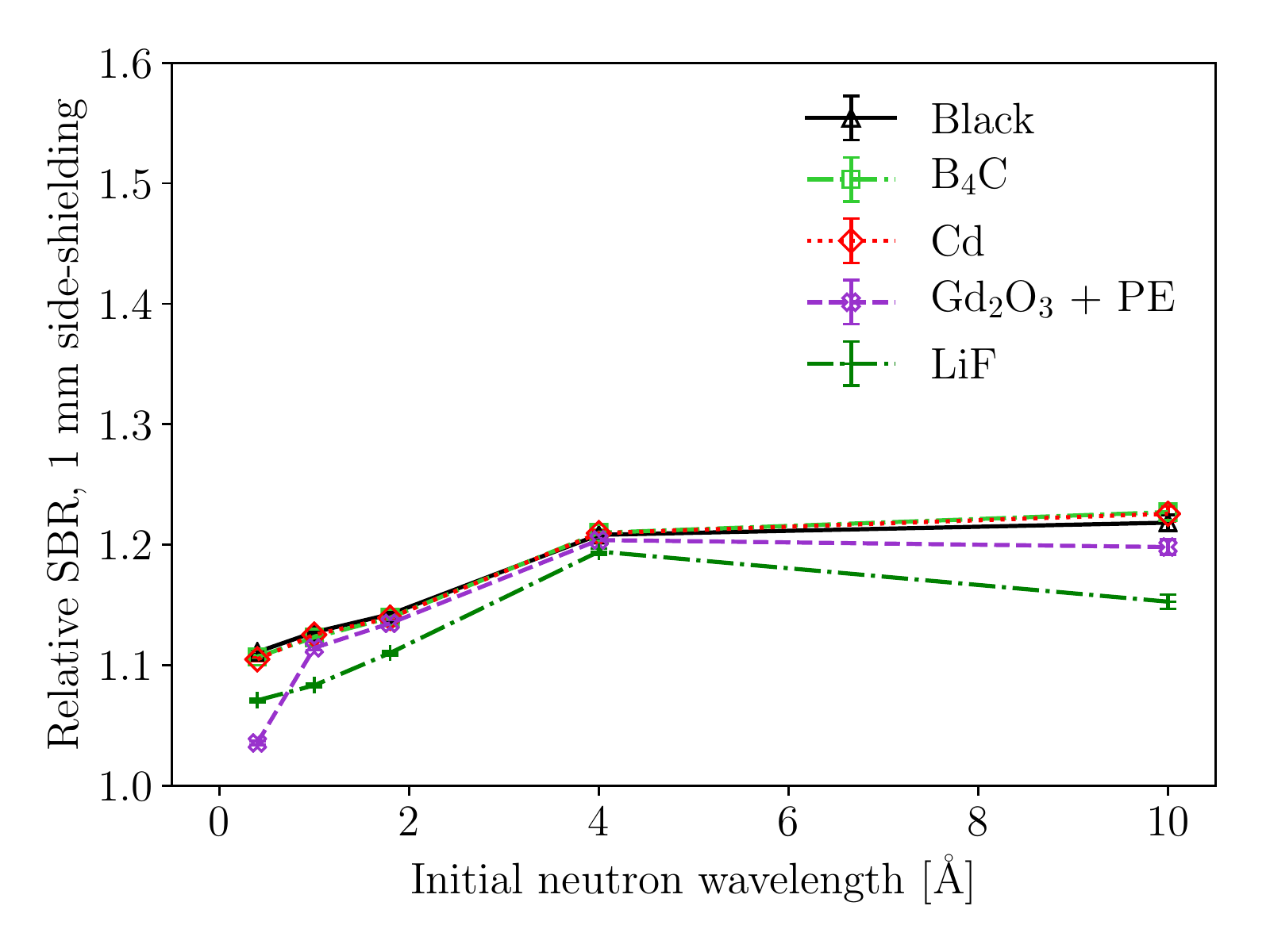}
    \caption{\label{MatIntVS}}
  \end{subfigure}

  \begin{subfigure}[b]{0.49\textwidth}
    \includegraphics[width=\textwidth]{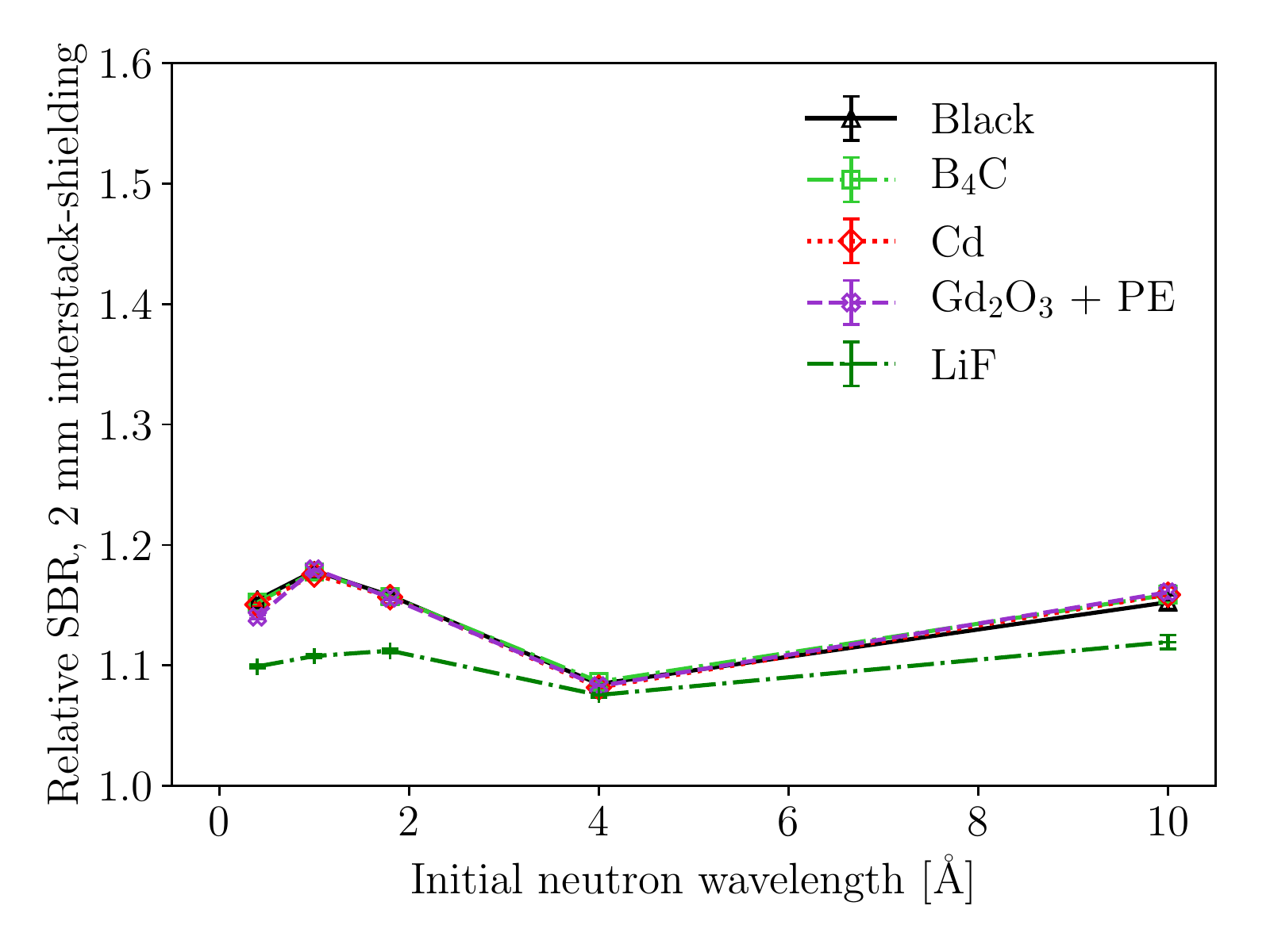}
    \caption{\label{MatIntStack}}
  \end{subfigure}
  
  \caption{\footnotesize Simulated Signal-to-Background Ratio (as specifically defined in equation~\ref{eq:SBR}) with different materials (including black material) for end-~(\subref{MatGridEnd}), side-~(\subref{MatIntVS}) and interstack-shielding~(\subref{MatIntStack}), normalised to the unshielded reference detector. 
    The statistical uncertainties are too small to be discernible.  \label{MatRelSBR}  }
  
\end{figure}

The effectiveness of a combined shielding for background suppression has been proven in the previous section (Section~\ref{black}). In order to obtain the best realisation of the ideal, combined shielding, common shielding materials (B$_{4}$C, Cd, Gd with polyethylene and LiF) are tested for the studied relevant shielding topologies. The impact of the different shielding materials are compared to the one of the black material at each shielding geometries in figure~\ref{MatRelSBR}.

It 
can be seen that 1~mm of B$_{4}$C or Cd as end-~(figure~\ref{MatGridEnd}) or side-shielding~(figure~\ref{MatIntVS}), and 2~mm of either of them as interstack-shielding~(figure~\ref{MatIntStack}), practically have equal background suppression capacity with the black material in the respective topologies. It is also shown, that the impact of the Gd-polyethylene mixture is also approximately the same as the one of the black material through the whole studied wavelength range, except at 0.4~\AA, where the impact of Gd is significantly lower in the case of the end-shielding and side-shielding. 
The SBR is increased with the application of end-shielding by 44-50\% in the case of black material, B$_{4}$C and Cd, and by 31\% in the case of the Gd with polyethylene. The same SBR increases for the side-shielding are 11\% and 4\%, respectively. On the one hand, these results confirm the effectiveness of Gd as shielding material at higher wavelengths, on the other hand they highlight the impact of the carrier media, especially the thermal scattering on the high hydrogen-content of typical carriers, like acrylic paint, glue, etc.

In figure~\ref{MatRelSBR} it is also shown that while the SBR can also significantly increase with the application of LiF shielding, this impact is much lower than the one of the previously discussed materials. The highest obtainable increase with LiF is 20\% at 0.4~\AA \ as end-shielding, and 15\% at 10.0~\AA \ as side-shielding. The respective quantities for black material are 50\% and 23\%. In essence, the impact of the different shielding materials is as expected on the basis of their respective cross sections.

\begin{figure}[t!]
  \centering
  \begin{subfigure}[b]{0.49\textwidth}
    \includegraphics[width=\textwidth]{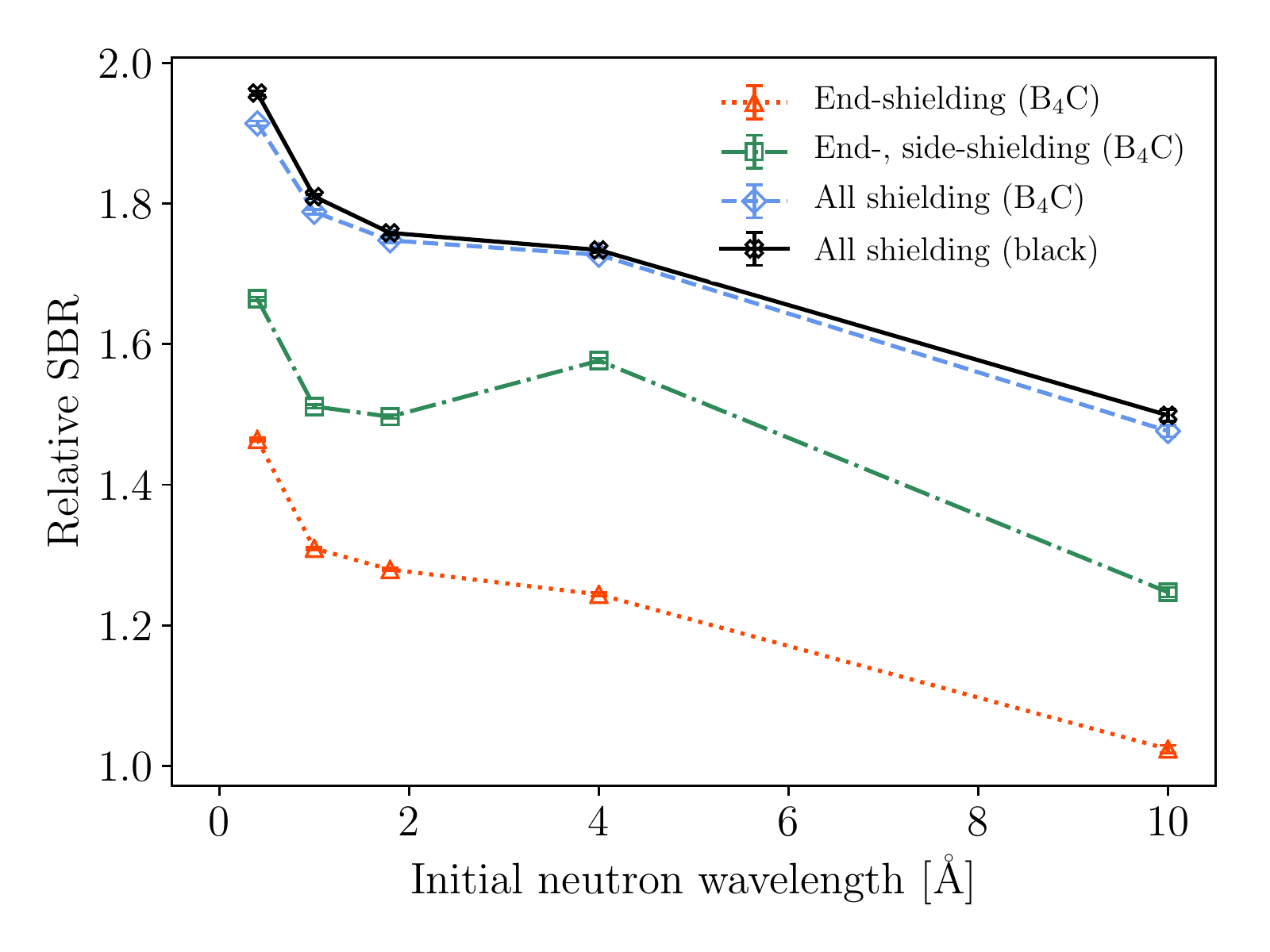}
    \caption{\label{combB4C}}
  \end{subfigure}
  \begin{subfigure}[b]{0.49\textwidth}
    \includegraphics[width=\textwidth]{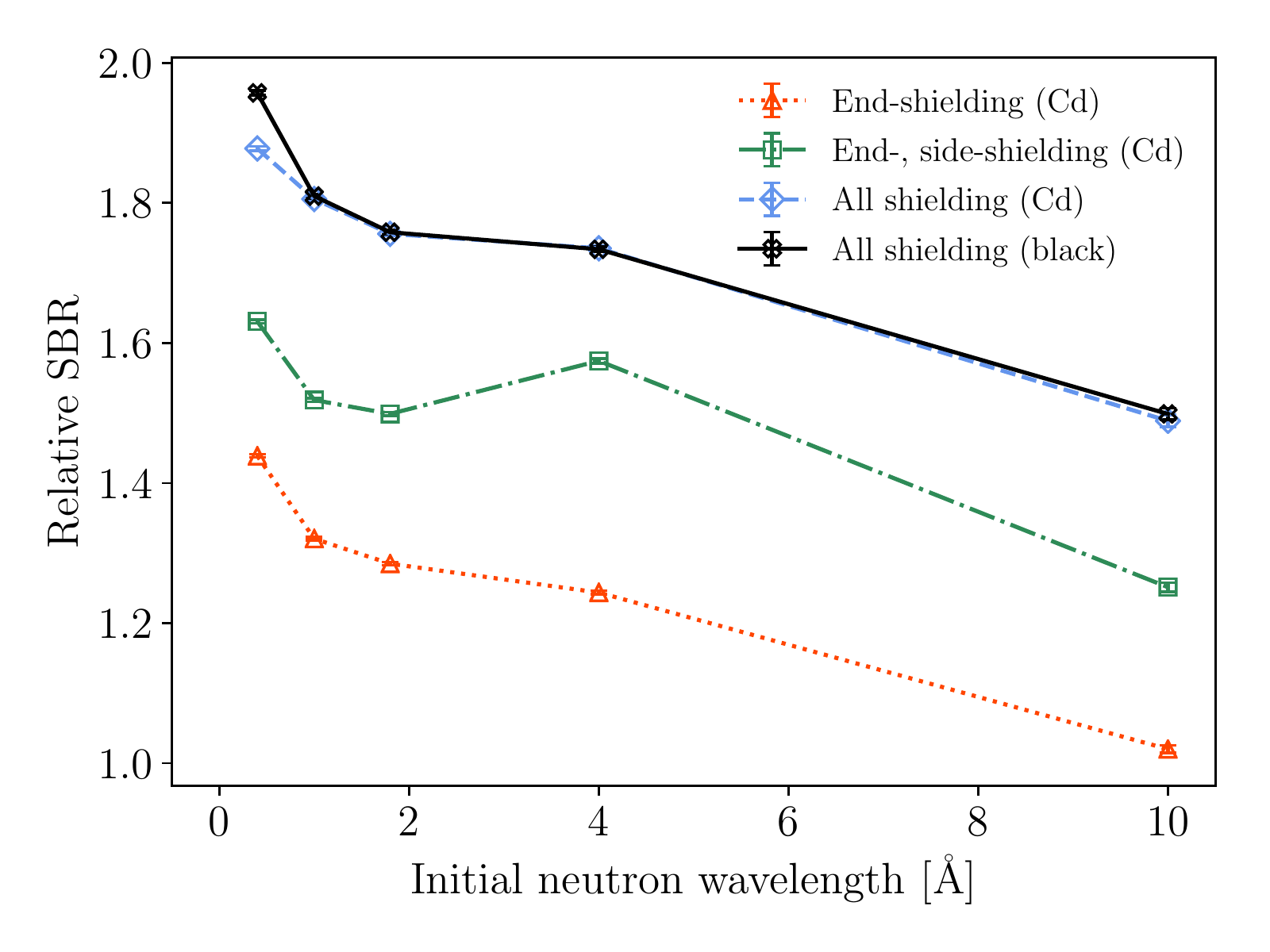}
    \caption{\label{combCd}}
  \end{subfigure}

  \begin{subfigure}[b]{0.75\textwidth}
    \includegraphics[width=\textwidth]{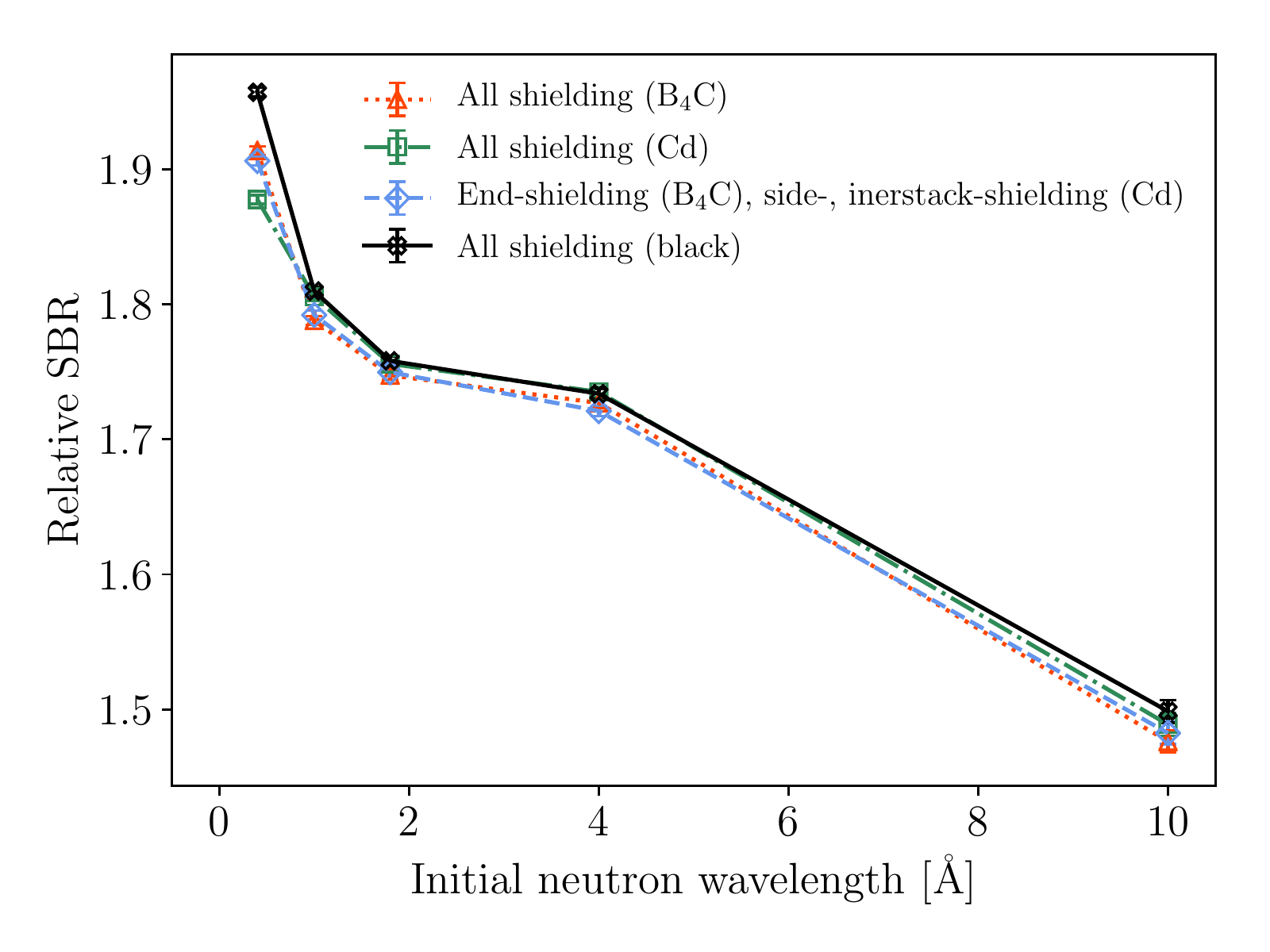}
    \caption{\label{combMix}}
  \end{subfigure}
  
  \caption{\footnotesize Simulated Signal-to-Background Ratio (as specifically defined in equation~\ref{eq:SBR}) with combined shielding with boron carbide~(\subref{combB4C}),  cadmium~(\subref{combB4C}), and both of them~(\subref{combMix}) compared to black material, normalised to the unshielded reference detector. 
    The statistical uncertainties are too small to be discernible.  \label{combSh}  }
  
\end{figure}

\begin{figure}[ht!]
  \centering
    \includegraphics[width=0.75\textwidth]{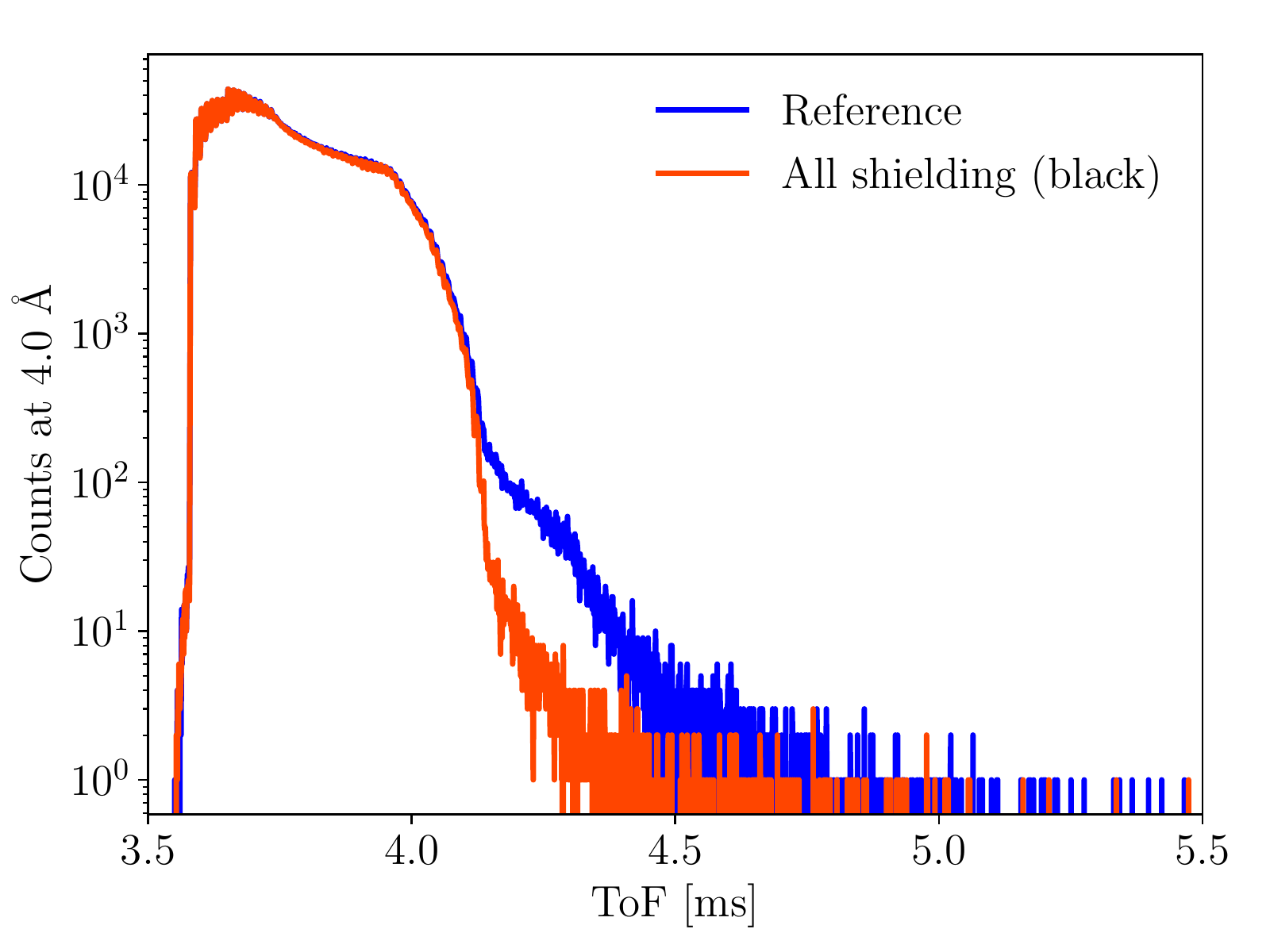}
  \caption{\footnotesize  Comparison of ToF spectra with and without shielding at 4~\AA~initial neutron wavelength. The overlapping curves do continue to the rear of the foreground curves. \label{ShToF}  }    
\end{figure}

As the B$_{4}$C and Cd are proven to be the appropriate 
realisation of black material in the CSPEC detector module, they are also applied in the afore-described complex shielding design. The realistic, complex shielding's impact on the SBR is shown and compared to the impact of black shielding in figure~\ref{combSh}. It is revealed, that 
whether B$_{4}$C~(figure~\ref{combB4C}) or Cd~(figure~\ref{combCd}) are applied for each shielding topologies, the total increase of SBR meets the one of the black material, as expected on the basis of results of the individually applied shielding geometries. However, it has to be 
noted, that the pure Cd shielding is less effective at 0.4~\AA. Here the obtained SBR increase is 96\% and 91\% with black absorber and B$_{4}$C, and 88\% with Cd, respectively. In figure~\ref{combMix} it is also shown, that this is the highest discrepance between the impact of the different shielding combinations, and that the differences between the respective SBR increases are within 1-2\% for all other wavelengths.
The beneficial effect of the combined shielding is also demonstrated in figure~\ref{ShToF} on the simulated ToF-spectrum: the background tail is cut down by one order-of-magnitude in the presence of the combined B$_{4}$C shielding.

Accordingly, with a realistic, B$_{4}$C and/or Cd based complex shielding design the SBR can be increased sufficiently close to the maximum theoretically obtainable value with the current operational parameters and design of the CSPEC detector module. The Gd is also proven to be a 
good shielding material, although the scattering on any carrier medium should be considered, especially at lower wavelengths.
In essence, common shielding materials are proven to be satisfactory for the CSPEC detector, and details of the complex shielding design can be chosen with regard to additional criteria, like cost, availability and engineering requirements.







\FloatBarrier
\section{Conclusion}\label{conc}

A novel, holistic approach is presented for shielding optimisation and background reduction in thermal and cold neutron detectors. 
The intrinsic scattered neutron background is determined for the Multi-Grid detector module of the CSPEC instrument at ESS. 

The effect of the long blade coating on the efficiency and Signal-to-Background Ratio is studied. It is revealed, that the efficiency can be increased by 8-19\%, and the SBR can be increased by 8-14\% in the 0.4-4.0~\AA \ 
wavelength region with the application of 1~$\mu$m $^{10}$B$_{4}$C coating on the long blades. The increase is 8\% and 13\% at the 4~\AA \ optimum of CSPEC, respectively. In terms of cost over neutron or SBR, the moderate increase in cost that can be expected by coating the long blades 
can be justified by 
the accompanying increase in SBR.

The contribution of the vessel and window on scattering is studied. It is shown that a decrease of SBR  with the increasing window thickness remains acceptable for a realistic, 1-5~mm thickness increase: 35\% maximum decrease with 5~mm thickness at 10~\AA, and <10\% decrease for all thicknesses at the 4~\AA \ optimum.  For this reason, the windows thickness can be chosen by engineering requirements.

The impact of  the aluminium vessel of the detector on the scattering is also determined, and proven to be equal or higher than the scattering on the window, pointing out the necessity of background suppression via internal detector shielding.

The background-reduction capacity of common shielding geometries, end-shielding, interstack-shielding and side-shielding are compared by applying black material. It is demonstrated, that the dominant shielding geometries are the end-shielding, absorbing 10-60\% of neutrons below 4~\AA, and the side-shielding, absorbing 5-10\% of neutrons through the whole wavelength range.

Common shielding materials, B$_{4}$C, Cd, Gd$_{2}$O$_{3}$ and LiF are tested for each shielding type, and 1~mm of B$_{4}$C or Cd is proven to be equally good shielding as the total absorber. With these materials as a combination of end-, side-  and interstack-shielding, the SBR can be raised by 91-50\% for 0.4-10~\AA~region, respectively.

\acknowledgments

This work has been supported by the In-Kind collaboration between the European Spallation Source (ESS~ERIC) and the Hungarian Academy of Sciences, Centre for Energy Research (MTA~EK). Richard Hall-Wilton, Kalliopi Kanaki, Anton Khaplanov and Thomas Kittelmann would like to acknowledge support from the EU Horizon2020 Brightness Grant~[grant number 676548]. Computing resources provided by DMSC Computing Centre (https://europeanspallationsource.se/data-management-software/computing-centre).









\end{document}